\documentclass[sigconf,balance=false]{acmart}
\usepackage{popets}
\setcopyright{popets}
\copyrightyear{YYYY}
\acmYear{YYYY}
\acmVolume{YYYY}
\acmNumber{X}
\acmDOI{XXXXXXX.XXXXXXX}
\acmISBN{}
\acmConference{Proceedings on Privacy Enhancing Technologies}
\usepackage{float}

\usepackage{graphicx}

\usepackage[frozencache,cachedir=minted-cache]{minted}

\usepackage{booktabs}

\usepackage{amsmath}
\usepackage{amsthm}

\theoremstyle{definition}
\newtheorem{definition}{Definition}
\newtheorem*{definition*}{Definition}

\usepackage{sparklines}
\usepackage[export]{adjustbox}
\usepackage{subfig}
\usepackage{xspace}
\usepackage{xcolor}
\usepackage{soul}
\usepackage{marginnote}
\usepackage{enumitem}
\usepackage[algoruled, linesnumbered, vlined]{algorithm2e}
\usepackage{setspace}
\usepackage{multirow}
\usepackage{multicol}
\usepackage{circledsteps}

\newcommand{\ie}{i.e.,\ }
\newcommand{\eg}{e.g.,\ }

\newcommand{\tinyskip}{\vspace{3pt}}
\newcommand{\mypar}[1]{\tinyskip\noindent\textbf{#1.}\xspace}

\newenvironment{myitemize}{%
\begin{itemize}[leftmargin=1em, itemsep=.1em, parsep=.1em, topsep=.1em,
    partopsep=.1em]}
{\end{itemize}}

\newenvironment{myenumerate}{%
\begin{enumerate}[leftmargin=1em, itemsep=.1em, parsep=.1em, topsep=.1em,
    partopsep=.1em]}
{\end{enumerate}}

\newenvironment{structure*}{\color{blue}\begin{myenumerate}}{\end{myenumerate}}

\newcommand{\update}[2][]{#2}
\newenvironment{updateblock}[1][]{}{}

\begin{document}

\title{Enabling Personal Dataflow Sovereignty via Bolt-on Data Escrow}


\author{Zhiru Zhu}
\orcid{0009-0007-7893-9017}
\affiliation{%
  \institution{The University of Chicago}
  \city{Chicago} 
  \state{IL} 
  \country{USA} 
}
\email{zhiru@uchicago.edu}

\author{Raul Castro Fernandez}
\orcid{0000-0001-7675-6080}
\affiliation{%
  \institution{The University of Chicago}
  \city{Chicago} 
  \state{IL} 
  \country{USA} 
}
\email{raulcf@uchicago.edu}



\begin{abstract}

    The digital economy is powered by a continuous and massive exchange of personal data. Individuals provide data to platforms in return for services, from social networking and search to health monitoring, entertainment, and access to LLMs. This exchange has created immense value, but it has also established a fundamental asymmetry of power: individuals possess only coarse-grained control over data access rather than fine-grained control over its purpose of use, creating a gap where data can be repurposed for undisclosed uses, \eg platforms selling the data to data brokers, which results in a critical loss of \emph{personal data sovereignty}. 
    
    This paper frames this socio-technical challenge as a \emph{dataflow} management problem. We propose a \emph{bolt-on data escrow} architecture through \emph{delegated computation}. In our model, instead of data flowing to platforms, platforms delegate their computation to a trustworthy escrow. This inversion empowers individuals with transparency and control over their dataflows. We present four contributions: (1) a dataflow model that explicitly incorporates computational purpose as a first-class primitive; (2) a minimally invasive programming interface, \textsc{run(access(), compute())}, built on a unified relational interface that virtualizes on-device data sources and a computation offloading component; (3) a concrete implementation of our escrow within the Apple ecosystem, demonstrating its practicality; and (4) both qualitative and quantitative evaluations demonstrating that our solution is expressive enough to implement a wide range of dataflows from real-world applications and introduces minimal runtime overhead. In summary, our work serves as a stepping stone toward achieving \emph{personal dataflow sovereignty}.
\end{abstract}


\maketitle

\section{Introduction}

Modern digital life runs on personal data. Every day, billions of individuals exchange sensitive data with platforms to enable services such as search, social media, health monitoring, entertainment, and access to LLMs. The prevailing model is simple: applications (\emph{apps}) pull sensitive data (\eg photos, contacts, location, health records) off individuals' devices and process the data on the platform's infrastructure. The moment a \emph{dataflow} crosses the individual's \emph{trust zone}---from the individual and their device to a remote platform---transparency and enforcement over \emph{how} the data will be used are lost. While existing permission systems gate access to raw data, they fail to bind or expose the \emph{purpose of its use}. This gap explains why seemingly benign requests (\eg ``share location for weather'') can covertly support unrelated uses such as profiling or resale.

We frame this as a data management problem. Specifically, a \emph{dataflow} management problem. We model a dataflow as a tuple $\langle A_s, D_s, f(), D_t, A_t\rangle$ that moves data $D_s$ from a source agent $A_s$ to a target agent $A_t$ via computation $f()$ that produces $D_t$. This makes the \emph{purpose} $f()$ explicit and enables \emph{dataflow preferences} that subsume traditional privacy preferences: individuals want control not just over \emph{what} data is shared, but \emph{with whom and for what purpose}, as well as where and how that computation runs based on preferences over performance, battery, cost, and others.

\mypar{Key idea: Bolt-on Data Escrow via Delegated Computation} We invert the prevailing model: instead of sending individuals' data to platforms, platforms \emph{delegate} their computation to a trustworthy intermediary, a \emph{data escrow} that resides within the individual's \emph{trust zone}. App developers specify their computation using a minimally invasive programming interface, \textsc{run(access(), compute())}. The \textsc{access()} function is built on top of a \emph{unified relational interface} that virtualizes on-device data sources containing individuals' sensitive data, allowing developers to declaratively specify their data needs using standard SQL. The \textsc{compute()} function contains the concrete implementation of the computation $f()$. The escrow executes both functions and returns only the result $D_t$. Developers never gain access to the raw $D_s$, and every dataflow is explicit and auditable by construction. The same interface also supports \emph{offloading} \textsc{compute()} across devices and escrow-controlled servers—allowing principled trade-offs between privacy, performance, battery, and cost.

\mypar{Why the Escrow is Deployable} Modern app ecosystems already concentrate data access behind Software Development Kits (SDKs)~\cite{koch2025impact, zhang2024navigating, rodriguez2025privacy}. \update[MR1, MR2, MR3, MR5]{We design an escrow that is \emph{bolted onto} Apple SDKs to enable practical path toward deployment. Apple can enforce the escrow as the bottleneck for data access through either its App Review process~\cite{app-review} or OS-level interventions. Specifically, by leveraging its existing enforcement mechanisms, such as entitlements~\cite{apple-entitlements}, Apple can ensure that any app attempting to bypass the \textsc{run()} interface to access sensitive data directly is either rejected during app review or terminated at runtime. The bolt-on design ensures that the only parties that see the change are developers, who already constantly adapt to mandatory SDK changes.} Unlike clean-slate solutions that require total ecosystem rewrites, our design integrates into the existing SDKs' update lifecycle. 

\mypar{Scope and limits} First, the main scope of our work focuses on establishing the \emph{mechanism} (the escrow) required to enforce dataflow preferences. While navigating how individuals practically define and interact with these preferences (\ie the \emph{policy}) remains an open human-computer interaction challenge, addressing it requires a viable escrow architecture to exist first.  \update[MR6]{Second, the escrow supports dataflows involving a single individual's data. If a computation calls third-party APIs, the escrow also makes this dataflow transparent (what leaves the device and why). However, fully supporting computation that combines multiple individuals' data or platform-confidential computation that cannot be disclosed remains future work. In these cases, our current model still improves the status quo by forcing platforms to use the escrow to implement a data copy that is subject to individuals' explicit approval.}

\mypar{Evaluation} We conduct a comprehensive evaluation that addresses three research questions regarding the escrow's feasibility, efficiency, and design. First, we perform a qualitative analysis, porting the dataflows of 10 popular, open-source iOS applications (\eg Wikipedia\cite{Wikipedia}, DuckDuckGo~\cite{DuckDuckGo}) to our escrow model to demonstrate whether the escrow's programming interface is expressive enough for real-world use. Our analysis shows that most dataflows can be effectively reimplemented using the escrow. Second, to demonstrate the practicality of the escrow, we conduct a quantitative evaluation by comparing the runtime of the escrow-based app with a baseline app that uses native Apple SDKs directly. Our results show that the escrow introduces very little overhead, due to the efficiency of the escrow's relational engine implementation. Finally, we evaluate three different strategies for implementing the escrow's relational engine to identify the performance trade-offs. We found that by implementing classic database query optimization techniques such as predicate pushdown, one of our strategies can achieve superior performance while still preserving data freshness. Together, these experiments provide convincing evidence that our data escrow is not only expressive enough to capture the complex dataflows of real-world applications but also introduces negligible overhead, making it a practical and deployable solution for enabling individuals to gain transparency and control over their dataflows.

\mypar{Contributions} This paper makes the following contributions:

\begin{myitemize}
  \item A \emph{dataflow model} that elevates the \emph{purpose} ($f()$) and \emph{trust zones}, enabling purpose-based, resource-aware dataflow preferences.
  \item A \emph{data escrow architecture} centered on a programming interface, \textsc{run(access(), compute())}, built on top of a unified relational interface and a computation offloading component.
  \item \update[MR1, MR2, MR5]{An \emph{escrow design and  implementation} in the Apple ecosystem. First, we map out the design space of the escrow and propose enforcement mechanisms to make the escrow's use mandatory for each design. We then explain the design of the escrow's relational engine; specifically, the relational schema design principles and three implementation strategies (Materialized Tables, Virtual Tables, and Virtual Tables with Pushdown).} Finally, we provide detailed technical descriptions of the escrow's relational engine implementation using Virtual Tables with Pushdown and its end-to-end computation offloading pipeline.
  \item Both \emph{qualitative and quantitative evaluation} showing that the escrow is expressive to implement dataflows from a wide range of real-world apps and introduces little performance overhead.
\end{myitemize}

\mypar{Outline} The rest of the paper is organized as follows. Section~\ref{sec:dataflow_model} introduces the dataflow model to reason about personal dataflow sovereignty. Section~\ref{sec:data_escrow_arch} presents the data escrow architecture and its programming interface. Section~\ref{sec:escrow_impl} describes the design of our escrow in the Apple ecosystem. Section~\ref{sec:implementation} details the technical aspects of our escrow implementation. Section~\ref{sec:evaluation} evaluates the escrow's expressiveness and performance. Section~\ref{sec:relatedworks} discusses the related work, and Section~\ref{sec:conclusion} concludes.

\section{A Model for Personal Dataflow Sovereignty} \label{sec:dataflow_model}

\begin{figure*}[t]
    \centering
    \includegraphics[width=\linewidth]{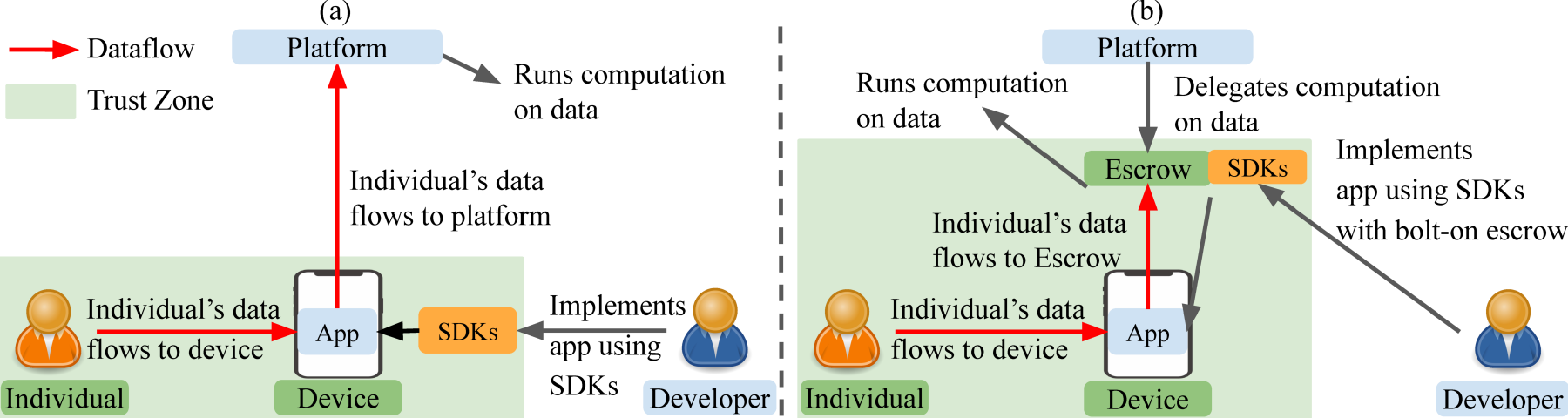}
    \caption{Dataflows in today's app ecosystem vs with the data escrow intermediary}
    \label{fig:dataflow}
\end{figure*}

To systematically address the problem of dataflow control, this section develops a model that allows us to reason about the flow of data between different entities. We begin by defining the primary \emph{agents} and the concept of a \emph{trust zone} in the app ecosystem (Section~\ref{sec:agent_trust_zones}). We then introduce the dataflow as the core unit of analysis (Section~\ref{sec:dataflow_def}). Building on this, we define dataflow preferences as a richer form of control that subsumes traditional privacy preferences (Section~\ref{sec:dataflow_pref}). Finally, we use this model to articulate the fundamental transparency and control deficits in the current app ecosystem that our work aims to solve (Section~\ref{sec:trans_control_deficit}).

\subsection{Agents and Trust Zones} \label{sec:agent_trust_zones}

We define an \emph{agent} as any entity with the ability to store and process data. There are many such agents in today's app ecosystem. We offer a distilled view of the ecosystem that highlights the agents that play a major role in the problem outlined in the introduction and the solution we present in this paper.

\begin{myitemize}
    \item The \textbf{Individual} is the data subject, the person whose personal data is at the center of these interactions.

    \item The \textbf{Device} is the hardware owned and operated by the individual, such as a smartphone, wearable, or computer. It is the primary repository for much of the individual's sensitive data.

    \item The \textbf{Platform} is the entity providing a service, such as a social media company, a search engine, or a health application provider. Platforms are represented by the applications running on the device and the backend infrastructure they control.

    \item The \textbf{Developer} is the entity employed by the \textbf{Platform} to develop applications. 
\end{myitemize}

Central to our model is the concept of the \emph{trust zone}. We define an individual's trust zone as the set of agents that the individual permits to access their raw data without restriction. In the prevailing model, this zone typically encompasses the individual and their personal devices. A dataflow is said to \emph{cross the trust zone} when its target agent lies outside this boundary. For instance, when a user (Individual) interacts with a social media app such as Instagram on their iPhone, the iPhone (Device) is within the user's trust zone. However, Instagram is operated by Meta (Platform), which is outside the trust zone. When the app requests access to the user's location or photos and transmits that data to Meta's servers for processing, that dataflow crosses the trust boundary, moving out from a trusted domain. It is precisely these boundary-crossing dataflows that require more control.

Figure~\ref{fig:dataflow}(a) provides a visual representation of today's app ecosystem, showing how an individual's data flows from their device, across the trust zone, to the platform's infrastructure where it is processed for purposes that are often opaque to the individual.

\subsection{Defining and Characterizing Dataflows} \label{sec:dataflow_def}

Individuals share their data with platforms, inducing data to flow. We model such flow with a \emph{dataflow}. 

\begin{definition}[Dataflow]
    \emph{A dataflow is a tuple $<A_s, D_s, f(), D_t, A_t>$, that indicates how the data $D_s$ from a source agent $A_s$ is transformed via function $f()$ to produce data $D_t$, which is accessible to a different target agent $A_t$.}
\end{definition}

A dataflow captures \emph{what} data is shared ($D_s$), with whom ($A_t$), and for what purpose\footnote{Technically, $f()$ limits the purposes for which the data can be used, rather than exactly determining the purpose, but the difference is not important for the technical presentation of this paper.} ($f()$). Using this model, we can characterize various common operations. A simple data copy, such as uploading a contact list to a platform's server, is a dataflow where $f()$ is the identity function and thus $D_s = D_t$. A more complex operation, like a weather app using an individual's precise location ($D_s$) to retrieve the local forecast ($D_t$), involves a dataflow where $f()$ is the API call to the weather service. The key insight is that the function $f()$ explicitly defines the purpose of the data use.

We are primarily interested in dataflows that involve individuals' sensitive personal data such as contacts, health, photos, and location data. While this data is used for purposes many individuals consider positive, such as sharing content with friends on social media or receiving localized weather forecasts, it has also been at the center of widespread privacy violations over the past two decades. Numerous high-profile incidents have revealed how platforms repurpose this data for undisclosed purposes: selling it to data brokers~\cite{crain2018limits}, enabling targeted surveillance~\cite{collier2022facebook}, training AI models without consent~\cite{yang2024creative}, or enabling discriminatory practices in housing and employment~\cite{spinks2019contemporary}. The opacity of these dataflows, where individuals cannot verify that their location shared for weather updates isn't also sold to advertisers, or that their health data isn't used to adjust insurance premiums, represents a critical failure of the current ecosystem to respect individual preferences and societal values around data use.

\subsection{Dataflow Preferences} \label{sec:dataflow_pref}

Individuals have preferences on how their data should be used, which go beyond privacy preferences. By making the purpose $f()$ an explicit part of our model, we define a richer and more powerful form of individual preferences that \emph{subsumes} traditional privacy preferences. Standard privacy preferences as presented by apps today are often binary, revolving around access to data (\eg "Allow access to Location?")~\cite{almuhimedi2015your, tan2014effect, felt2012android}. \emph{Dataflow preferences}, in contrast, enable nuanced, purpose-based control over individuals' data.

We argue that an individual's privacy is preserved only when individuals' dataflow preferences are preserved. An individual may be perfectly willing to permit a dataflow where their location data ($D_s$) is used by a weather app ($A_t$) for the purpose of fetching a forecast ($f()$). However, they may wish to deny a different dataflow where the same location data is sent to an advertising broker ($A_t$) for the purpose of profiling ($f()$). This aligns with the privacy framework of Contextual Integrity~\cite{nissenbaum2004privacy}, which states that adequate privacy protection can only be offered when information norms of a specific context are respected. Our dataflow model provides the formal structure needed to define and enforce these norms. It allows individuals to express preferences not just on
what data is shared, but for what purpose, which is a more accurate reflection of real-world privacy expectations.

Furthermore, dataflow preferences extend beyond traditional privacy preferences. For instance, an individual might specify a preference to only allow computationally intensive dataflows (\eg video processing) to run when their device is connected to Wi-Fi and charging, in order to conserve battery life and mobile data. Another individual might prefer that certain computations are offloaded to a server for faster performance. These dataflow preferences concerning resource consumption, performance, and cost are not related to privacy in the classic sense, but they are critical aspects of controlling one's data use. By capturing a broad set of dataflow preferences, our model allows us to reason about \emph{personal dataflow sovereignty}---individuals' fundamental right to control their dataflows is the right to enforce their dataflow preferences.

\subsection{The Transparency and Control Deficits} \label{sec:trans_control_deficit}

The dataflow model allows us to articulate two key challenges that result in the loss of \emph{personal dataflow sovereignty}:

\begin{myitemize}
    \item \textbf{Lack of Transparency}: Individuals do not have clear visibility on what dataflows take place when using an app. When an app requests their data, the purpose $f()$ is at best implied and at worst intentionally obscured~\cite{almuhimedi2015your, zang2015knows, claesson2020technical, xiao2023lalaine, rodriguez2024sharing}. The terms of service that supposedly govern data use are notoriously broad and vague, failing to specify the concrete computations that will be performed on the data~\cite{obar2020biggest}.

    \item \textbf{Lack of Control}: Even if the purpose $f()$ were made transparent, individuals currently lack the technical mechanisms to enforce their preferences on the dataflows. The only available control is a blunt instrument: denying access to the source data $D_s$ entirely. This often forces an undesirable trade-off, as denying access may break the primary, desired functionality of the app, leaving the individual with an all-or-nothing choice.
\end{myitemize}

\noindent\textbf{The goal of this paper} is to design and implement the technical \emph{mechanism} that endows individuals with both \emph{transparency} and \emph{control} over their dataflows, providing the necessary foundation for future \emph{policy} design on eliciting individuals' dataflow preferences.

\section{The Data Escrow Architecture} \label{sec:data_escrow_arch}

To address the transparency and control deficits identified in the current app ecosystem, this section details the design of our proposed \emph{data escrow} architecture. We first introduce our key idea of employing the delegated computation model (Section~\ref{sec:delegated}). We then present a minimally invasive programming interface to enable delegated computation (Section~\ref{sec:programming_int}). Next, we introduce computation offloading as a core architectural feature that enables moving computation across devices (Section~\ref{sec:offloading}). We then describe the data virtualization layer that provides a unified relational interface for transparent data access (Section~\ref{sec:data_virt}). Finally, we provide a taxonomy of common dataflow patterns to clarify the capabilities and limitations of our solution (Section~\ref{sec:taxonomy_dataflow}).

\subsection{Key Idea: Delegated Computation Model} \label{sec:delegated}


The key idea of our solution is to replace the traditional model of sending individuals' data to the platform, illustrated in Figure~\ref{fig:dataflow}(a), with a \emph{delegated computation} model. This architectural shift is visualized in Figure~\ref{fig:dataflow}(b). Instead of the platform pulling an individual's data ($D_s$) from their device to run computation ($f()$) on its own opaque infrastructure, which crosses the individual's trust zone, the platform must delegate its computation to a trustworthy \emph{data escrow}~\cite{xia2023data} intermediary (\ie it must be in the individual's trust zone). This inversion of control is made possible because modern applications are not built from scratch; they are built using \emph{Software Development Kits (SDKs)} provided by the ecosystem owner (\eg Apple or Android). These SDKs provide the standard tools and APIs for all essential functions, including access to sensitive personal data. Our key insight is that an escrow can be \emph{bolted onto} these existing SDKs. Developers, who already rely on these SDKs to write apps, would now use an escrow-based programming interface to specify their computation.

As shown in Figure~\ref{fig:dataflow}(b), the developer implements their app against this new programming interface, effectively delegating their intended computation, $f()$, to the escrow. The escrow, operating as a trustworthy intermediary, then pulls the required data ($D_s$) from the device and executes the $f()$ on behalf of the platform. This execution only proceeds if the individual approves the entire dataflow. Because the escrow mediates all access to sensitive data, it becomes a natural \emph{bottleneck}, making every dataflow inherently transparent and controllable by design. The fundamental trust assumption shifts: the individual no longer needs to trust every application platform but instead places their trust in the escrow provider.

\begin{table*}[ht!]
\centering
\resizebox{\textwidth}{!}{%
\begin{tabular}{|c|c|c|c|c|}
\hline
Pattern &
  Within individual's trust zone &
  Computation with external API call &
  Cross-individual computation &
  Confidential computation \\ \hline
Description &
  \begin{tabular}[c]{@{}c@{}}Computation runs fully \\ within trust zone using a \\ single individual's data\end{tabular} &
  \begin{tabular}[c]{@{}c@{}}Computation calls third-party or \\ platform API, sending individual's \\ data outside the trust zone\end{tabular} &
  \begin{tabular}[c]{@{}c@{}}Computation combines data \\ from multiple individuals to \\ produce aggregate results\end{tabular} &
  \begin{tabular}[c]{@{}c@{}}Platform cannot or will not \\ reveal computation\end{tabular} \\ \hline
Example &
  \begin{tabular}[c]{@{}c@{}}An app scans individual's photo \\ library to tag pictures of their \\ pets using image classification \\ model hosted on-device or in \\ an escrow-controlled server\end{tabular} &
  \begin{tabular}[c]{@{}c@{}}A weather app gets individual's\\ location, calculates an approximate\\ region based on the location, and\\ sends the region to a third-party \\ weather API to fetch forecast\end{tabular} &
  \begin{tabular}[c]{@{}c@{}}A mapping service using \\ location data from multiple \\ individuals to calculate \\ real-time traffic conditions\end{tabular} &
  \begin{tabular}[c]{@{}c@{}}A social media app using a \\ proprietary algorithm to \\ recommend friends from \\ individual's contacts\end{tabular} \\ \hline
\begin{tabular}[c]{@{}c@{}}Supported\\ by escrow\end{tabular} &
  Fully supported &
  \begin{tabular}[c]{@{}c@{}}Supported with transparency - \\ The escrow makes data transfer\\  transparent\end{tabular} &
  \begin{tabular}[c]{@{}c@{}}\update{Currently supported by}\\\update{copying each individual's}\\\update{data to platform}\end{tabular} &
  \begin{tabular}[c]{@{}c@{}}\update{Currently supported by}\\\update{copying each individual's}\\\update{data to platform}\end{tabular} \\ \hline
\end{tabular}%
}
\caption{Taxonomy of dataflow patterns and whether they are supported by the escrow's delegated computation model}
\label{tab:dataflow_pattern}
\end{table*}

\subsection{The \textsc{run(access(), compute())} Interface} \label{sec:programming_int}

\update{To enable delegated computation, we design a minimally invasive programming interface centered on a higher-order function: \textsc{run(access(), compute())}, which has two key parameters:}

\begin{myitemize}
    \item \textsc{access()}: The \emph{data access function} that specifies the input data, $D_s$, required for the subsequent computation $f()$. As detailed in the next section, developers implement this function declaratively by writing a standard SQL query.

    \item \textsc{compute()}: The \emph{computation function} that contains the concrete implementation of $f()$. It takes the output of \textsc{access()} as its input, and it can use any libraries or call third-party APIs.
\end{myitemize}

The \textsc{run()} function is the core entry point executed by the escrow. When an application calls \textsc{run()}, the escrow orchestrates the entire dataflow. It first executes the \textsc{access()} function to retrieve $D_s$. \update[MR3]{It then executes the \textsc{compute()} function within an isolated execution environment managed by the escrow, passing $D_s$ as the input, to produce the result $D_t$.} The critical guarantee of this interface is that the platform and its developers never gain access to the raw data $D_s$. They provide the code for \textsc{access()} and \textsc{compute()}, and they receive the final output $D_t$, but $D_s$ is only ever handled by the escrow within the individual's trust zone. This cleanly separates the specification of computation from the access to raw data.

\mypar{Delegated Computation in Action} To make the delegated computation model concrete, consider a simple weather app. The app's goal is to use the individual's current location ($D_s$) to fetch the local weather forecast ($D_t$). Without the escrow, a developer would use the native SDK to directly fetch the device's location coordinates, which are considered sensitive, then send those coordinates across the individual's trust zone to the platform's server. A simplified code snippet (in Apple's Swift language) might look like this:

\begin{minted}{Swift}
// Use native SDK to fetch raw location coordinates
let location = LocationManager.getCurrentLocation()
// Send raw data across the trust zone to the platform
let weather = PlatformAPI.fetchWeather(for: location) 
\end{minted}

In this common pattern, the raw location coordinates leave the individual's device. The individual has no technical guarantee that the platform is only using the data to fetch the weather and not for other purposes, such as selling it to data brokers. \update{Next, we show how the developer writes the app using the \textsc{run()} interface.} 

\begin{updateblock}
\begin{minted}{Swift}
import Escrow
// Declaratively specify the data needed
let locationSQL = "SELECT cllocation FROM Location  
                   ORDER BY timestamp DESC LIMIT 1"
// Define the computation (purpose)
func getWeather(location: CLLocation) -> Weather {
    // Only sends an approximate region to platform
    let region = getSurroundingRegion(from: location)
    return PlatformAPI.fetchWeather(for: region) }
// Delegate computation to escrow
let weather = Escrow.run(locationSQL, getWeather)
\end{minted}
\end{updateblock}

\update[MR3]{In this revised implementation, the app never handles the raw location data. Since the platform only needs the approximate region around the individual's current location to fetch the weather, the app only sends the approximate region to the platform. The location SQL query and the \texttt{getWeather} function are submitted to the escrow. The escrow executes the query, passes the resulting \texttt{CLLocation}~\cite{cllocation} object to the \texttt{getWeather} function, then computes and sends the approximate region to the platform to fetch weather, and finally returns only the \texttt{Weather} object to the app. The entire operation is transparent to the individual, who can verify the purpose and approve or deny this dataflow, with the understanding that their raw location coordinates never leave their trust zone. Alternatively, if the platform needs the raw location instead of the approximate region to fetch weather, the developer can still implement the same logic as before, which directly sends the raw data to the platform. This explicit data copy will be transparent to the individual via the escrow. Unlike the non-escrow version, where individuals do not know what happens to their data after they grant their permission for the app to access their location, they now have a clear idea that their raw location data is sent to the platform, and they can block the escrow from executing this data copy.}


\subsection{Computation Offloading} \label{sec:offloading}


The delegated computation model naturally extends to support flexible execution locations. While many computation can run efficiently on-device, some may be too resource-intensive or require access to server-side resources. The practice of computation offloading---transferring intensive tasks to an external location---is a well-established technique to overcome device limitations such as computational power, storage, and energy~\cite{kumar2013survey}. Our escrow architecture allows \textsc{compute()} functions to be offloaded to a trustworthy, \emph{escrow-controlled} server. This is achieved by extending the \textsc{run()} function with an optional \textsc{server\_id} parameter: \textsc{run(access(), compute(), \textsc{server\_id})}. 
When \textsc{server\_id} is provided, the on-device escrow first executes the \textsc{access()} function to obtain the input data $D_s$. It then serializes $D_s$ and transmits it, along with the signature of the \textsc{compute()} function to be executed, to the specified remote server via a secure channel. The remote server, which is managed by the escrow provider and thus remains within the individual's trust zone, executes the computation and returns the serialized result $D_t$. This offloading mechanism allows developers to balance on-device resources against the power of server-side processing. It aligns well with emerging industry trends such as Apple's Private Cloud Compute (PCC)~\cite{private-cloud-compute}, which provides a secure server infrastructure designed for privacy-preserving computation offloaded from mobile devices. Our escrow architecture provides a general framework that could be directly implemented on top of such infrastructure, leveraging its strong, hardware-enforced security guarantees.

\subsection{Data Virtualization via Relational Interface} \label{sec:data_virt}

\begin{figure}[h]
    \centering
    \includegraphics[width=\linewidth]{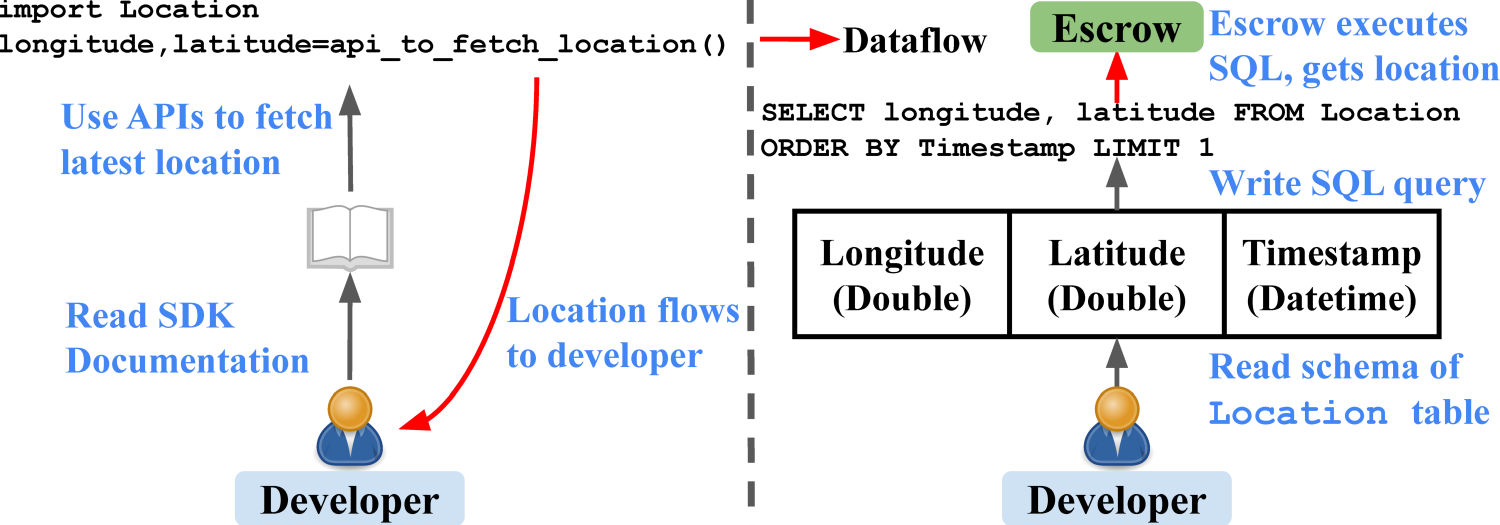}
    \caption{Accessing location using native SDKs (left) vs using the escrow's relational interface (right).}
    \label{fig:relational}
\end{figure}


A key design decision is how developers should implement the \textsc{access()} function. Simply allowing developers to write arbitrary code would make the data access function opaque, effectively hiding a secondary computation within the \textsc{access()} function itself. To ensure transparency, we design a \emph{declarative} interface based on the classic data virtualization approach~\cite{carey1995towards}.

The escrow virtualizes the device's heterogeneous data sources, such as contacts, photos, and location updates, into a unified relational schema. Instead of using numerous native SDK APIs to access data, developers now implement \textsc{access()} by writing a standard SQL query against this public schema. For example, to fetch an individual's most recent location, a developer would write an SQL query rather than importing the \texttt{Location} framework and writing imperative code to request location updates. Figure~\ref{fig:relational} contrasts these two approaches. The traditional approach (left) requires developers to learn SDK-specific APIs and gives them direct access to the resulting location. The escrow's relational interface (right) simplifies the data access to a declarative query and ensures the developer never handles the raw location data directly.

\subsection{Taxonomy of Dataflow Patterns} \label{sec:taxonomy_dataflow}

The delegated computation model is designed to give individuals control over a wide range of dataflows, but its applicability varies depending on the nature of the computation. To clarify the scope of the escrow, we provide a principled categorization of four common dataflow patterns in Table~\ref{tab:dataflow_pattern}. 

The escrow fully supports computations that are self-contained within an individual's trust zone, which includes computation on-device or in escrow-controlled servers through computation offloading. This category covers a wide range of common and important dataflows in today's apps. \update[MR6]{The escrow also supports computation that require calling external APIs, which makes the data transfer transparent. However, fully supporting dataflows that are cross-individual or confidential remains future work. For instance, computing real-time traffic conditions requires combining location data from many individuals, and platforms may be unable to delegate proprietary algorithms that are trade secrets. In those cases, the escrow still forces platforms to implement a data copy. This still represents a significant improvement over the status quo: the data copy becomes transparent and subject to the individual's explicit approval, empowering them with a meaningful choice that is often absent in today's ecosystem. Leveraging technologies such as Secure Multi-Party Computation (SMPC)~\cite{canetti1996adaptively} and Homomorphic Encryption (HE)~\cite{acar2018survey} offers a theoretical pathway to support these patterns within the delegated model. For instance, if mobile hardware and network bandwidth evolve to support the overhead of these protocols efficiently~\cite{gamiz2025challenges}, the escrow could orchestrate MPC dataflows where multiple devices jointly compute aggregate computation without revealing individuals' data, or evaluate HE-encrypted proprietary algorithms.}


\section{Escrow in Apple Ecosystem} \label{sec:escrow_impl}

To demonstrate the practicality of our proposed escrow architecture, we design and implement a data escrow within the Apple ecosystem. We begin by explaining our rationale for choosing the Apple ecosystem and how it provides a natural bottleneck for data access (Section~\ref{sec:exploit_sdk}). \update{We then describe mechanisms to make the escrow's use mandatory for each escrow design (Section~\ref{sec:escrow_integration}).} Finally, we describe our schema design principles and three implementation strategies of the escrow's relational engine (Section~\ref{sec:relational}). 

\subsection{Exploiting Apple SDKs as Bottleneck} \label{sec:exploit_sdk}

Mobile devices are the primary repositories of sensitive personal data, and the two dominant ecosystems are Apple's iOS and Google's Android~\cite{mobile-threat-report}. While our escrow architecture could be adapted to either, we choose to implement our prototype in the Apple ecosystem, with Apple acting as the trustworthy escrow, as its vertically integrated nature offers a clear path to deployment. A successful escrow must become the \emph{bottleneck} to mediate all sensitive data access. The Apple ecosystem facilitates the creation of such a bottleneck for three key reasons: \update[MR2]{\textbf{First}, today, to access sensitive personal data such as health records, photos, contacts, or location, developers are required to use specific Apple SDKs (\eg Contacts~\cite{swift-contacts}, PhotoKit~\cite{photokit}, CoreLocation~\cite{swift-core-location}). There are no alternative methods to access the data. \textbf{Second}, all apps distributed through Apple's App Store must undergo a rigorous review process before they can be published~\cite{app-review}. The App Review provides a centralized point of control to ensure that all data access in the app goes through the escrow.} \textbf{Third}, individuals generally exhibit a high degree of trust in their Apple devices and, by extension, in Apple as the platform vendor. This pre-existing trust makes the platform vendor a natural and suitable candidate to assume the role of the data escrow, as it would seamlessly fit within the individual's existing trust zone. Combining these factors makes it feasible to create an escrow that can be adapted to the existing Apple infrastructure. 

\mypar{\update[MR6]{Adapting escrow to Android ecosystem}} \update{Enforcing the escrow on Android presents distinct challenges due to Android's support for third-party app stores, Original Equipment Manufacturer (OEM) customizations (\ie companies such as Samsung and Xiaomi building customized Android operating systems on top of Google's standard Android and offering their dedicated app stores), and sideloading (\ie individuals can install apps that are not published from an official app store)~\cite{android-sideloading}. To adapt the escrow to the Android ecosystem, a centralized, trustworthy app distributor, such as Google Play or a dedicated privacy-focused app store, could mandate the integration of the escrow as a strict prerequisite for app distribution and review apps to ensure all data access goes through the escrow.}
\vspace{-0.1cm}

\subsection{\update{Escrow Design and Enforcement}}\label{sec:escrow_integration}

\update[MR1, MR2, MR3, MR5]{We discuss two design choices for the escrow: a library-based integration and an OS-level system service. Because modifying proprietary systems by Apple is infeasible for third-party researchers, our escrow prototype implements the library-based design. However, we detail both designs to demonstrate how Apple enforces the escrow as a non-bypassable bottleneck for data access and  how the escrow enforces control.}

\mypar{\update{Library-Based Design}} \update{In this design, the escrow is a compiled library (\eg an iOS framework) that developers link into their apps. Enforcement relies on Apple's App Review process, in which apps are reviewed and scanned for unauthorized API usage~\cite{app-review}; enforcing the escrow simply extends this process. Apps that attempt to call native data access SDKs directly, rather than declaratively describing data to access through the \textsc{access()} function in \textsc{RUN()}, are flagged and rejected during submission.}

\mypar{\update{OS-Level Service Design}} \update{An alternative design, which Apple could adopt, integrates the escrow directly into the operating system as a privileged service. In the Apple ecosystem, access to sensitive data is governed by the Transparency, Consent, and Control (TCC) mechanism~\cite{apple-tcc}. In this design, the OS uses its existing (TCC) mechanism to deprecate direct sensitive data entitlements~\cite{apple-entitlements} that apps require to access data. Only the privileged escrow service retains the entitlements to read sensitive data. If an app attempts to bypass the escrow and read protected resources directly from the filesystem or memory without the required entitlements, the OS kernel intercepts and terminates the process.}

\mypar{\update{Enforcing Control}} \update{Once the escrow becomes the bottleneck for data access, it must enforce control in two areas: (1) controlling data access so an app cannot gain access to data it did not explicitly request via the \textsc{access()} function in \textsc{run()}, and (2) ensuring the \textsc{compute()} function does not contain malicious logic that extracts or transfers data indirectly, which would jeopardize the escrow's transparency. Apple's existing enforcement mechanisms naturally address both concerns. First, to control data access, developers must explicitly declare their required capabilities to access any protected resource~\cite{apple-capabilities}, similar to how they do today. Depending on the integration level, this relies on standard entitlements (library-based design) or escrow-specific capabilities (OS-level design). If an app declares a capability (\eg accessing the Photos library) but lacks the corresponding \textsc{access()} function (\eg querying the escrow's \texttt{Photos} table), the app is flagged during App Review. Conversely, attempting to access data without the required capability results in runtime denial. Second, to prevent malicious data extraction and transfer, developers must still provide a Privacy Manifest~\cite{apple-privacy-manifest} that declares the data collected, its purpose, and any network domains used for data transfer. During Apple's mandatory App Review process~\cite{app-review}, reviewers evaluate the app's \textsc{run()} implementation against these declarations. If the computation contains malicious logic to indirectly extract data or communicates with undeclared domains, the app is rejected prior to publication, ensuring the escrow's transparency remains uncompromised.}

\subsection{Relational Interface Design} \label{sec:relational}

\update{The core component of the escrow is the data virtualization layer that provides the relational interface for the \textsc{access()} function. We detail the relational interface design of our escrow prototype to demonstrate a practical path toward deployment.}

\subsubsection{Designing the Relational Schema} The escrow exposes a public relational schema that developers use to formulate their data access queries. The design of this schema is guided by the principle of minimizing friction for developers who are already familiar with the native SDKs. Rather than creating an entirely abstract schema, our tables and columns map closely to the objects and properties provided by Apple's SDKs~\footnote{Nevertheless, the \emph{object–relational impedance mismatch}~\cite{ireland2009classification} always exists. Our goal is not to solve this mismatch, but to design the schema in a developer-friendly manner.}. For example, Apple's \texttt{PhotoKit}~\cite{photokit} framework represents a photo or video asset as a \texttt{PHAsset}~\cite{phasset} object. Our corresponding
\texttt{Photos} table in the relational schema includes columns for standard attributes such as \texttt{id} (\texttt{String}), \texttt{mediaType} (\texttt{Integer}), and \texttt{creationDate} (\texttt{Date}). Crucially, it also includes a \texttt{phasset} column whose type is the native \texttt{PHAsset} object itself. This allows a developer to write a SQL query to filter and select assets declaratively, and then receive a sequence of native \texttt{PHAsset} objects in the \textsc{compute()} function, on which they can perform operations using the standard \texttt{PhotoKit} APIs they already know. This design choice makes the transition to the escrow model more intuitive and less disruptive for developers.

\subsubsection{Designing the Relational Engine} The relational interface provides a logical view of the data to developers, and they no longer need to implement the concrete mechanisms to retrieve the data---this responsibility falls onto the escrow. We design, implement, and evaluate three approaches to implement the escrow's relational engine, each with different performance trade-offs. \label{sec:relational_stratedgies}

\mypar{Materialized Tables} In this baseline approach, the escrow upon initialization fetches all relevant data declared by the app's \textsc{access()} functions using the native Apple SDKs, and materializes it into tables within an on-device SQLite database. The \textsc{access()} function is then executed as a standard SQL query against this database. While this approach benefits from the performance of a native database engine, it suffers from two significant drawbacks. First, it creates a data freshness problem, as the materialized data can become stale if the underlying source changes. Second, it introduces significant overhead for complex data objects like \texttt{PHAsset}, which cannot be stored directly in SQLite. Instead, we must store a reference ID, and the escrow must perform a costly post-query lookup to map these IDs back to their corresponding native objects.

\mypar{Virtual Tables} To address the issues of data freshness and object handling, our second approach leverages SQLite's virtual table mechanism~\cite{sqlite-vtab}. A virtual table is a schema object that looks like a normal table but does not store any data on disk. Instead, when the database engine needs to scan a virtual table, it makes callbacks to custom C functions provided by our implementation. We implement these callbacks to make live API calls to the relevant Apple SDKs (\eg \texttt{PhotoKit}, \texttt{Contacts}) to fetch all relevant data needed to populate the table. This data is then passed back to the SQLite engine, which performs any necessary filtering, sorting, or aggregation in memory. This approach ensures that query results are always up-to-date and avoids data duplication. Furthermore, because the data is only held in memory during query execution, we can pass pointers to complex native Swift objects (like \texttt{PHAsset}) through the C-to-Swift bridge, allowing the \textsc{compute()} function to receive them in their native format without any serialization overhead. However, this approach can be inefficient for selective queries, as it requires fetching all data from an API before filtering.

\mypar{Virtual Tables with Pushdown} Our third approach enhances the virtual table implementation by applying classic database query optimization principles. The goal of predicate pushdown~\cite{hellerstein1993predicate} is to filter data as early as possible to minimize data processing. We implement logic to push down query operations such as projections, predicates, sorting, and limits, from the SQLite query engine into the underlying native SDK API calls, since many of Apple's SDKs provide APIs that allow for more efficient, constrained data fetching. For example, consider the query \texttt{SELECT phoneNumber FROM Contacts WHERE givenName == 'Alice'}. A naive virtual table implementation would fetch \emph{all} contact records from the device and let SQLite perform both the filtering on \texttt{givenName} and projection on \texttt{phoneNumber}. Our pushdown implementation, however, will translate this query into an efficient \texttt{CNContactFetchRequest}~\cite{cncontactfetchrequest} with a predicate for the name 'Alice' and \texttt{keysToFetch} set to only the \texttt{phoneNumber}. This pushes down the filtering and projection operations from the SQLite engine into the native API, reducing the amount of data that needs to be fetched from the underlying data source and processed by the SQLite engine, leading to significant performance gains in many cases, as shown in our evaluation.


\subsection{\update{Bootstrapping Dataflow Preferences}}
\update[MR4]{While the primary contribution of our work is the technical mechanism required to enforce dataflow preferences (the escrow), an effective system must also practically elicit these preferences from individuals. We propose a concrete design that piggybacks on Apple's existing permission infrastructure to demonstrate a technically viable mechanism for purpose-bound control.

Currently, Apple elicits individuals' privacy preferences through its TCC subsystem. When an app requests sensitive data, TCC triggers a system-rendered pop-up (\eg "App X would like to access your photos") alongside a developer-provided purpose string explaining why the app needs access~\cite{apple-purpose-string}; developers are required to provide useful purpose strings so individuals using the app can decide whether to allow or deny the app's access to their data, and App Review will reject apps that access protected resource without providing a purpose string~\cite{app-review}. Our design builds directly upon this existing practice. In the escrow model, the mandatory purpose string is elevated: instead of providing arbitrary explanations of why the app needs access to certain data, developers are required to provide a specific purpose string that directly maps to the logic of the \textsc{compute()} function for every \textsc{run()} invocation. During the App Review process, the reviewers verify that this declared purpose string accurately reflects the \textsc{compute()} logic; this is an extension of their existing review protocol. At runtime, the OS intercepts the \textsc{run()} call and generates an extended TCC prompt. Instead of a generic data request, this prompt displays the requested source data ($D_s$), the target agent ($D_t$) if the data leaves the device, and the vetted purpose string explaining the purpose ($f()$). This integration bridges the escrow architecture with Apple's existing UI, providing individuals with transparency and control over their dataflows without revamping the App Review process.

However, we explicitly note the limitations of this design. Prompting an individual for every distinct \textsc{run()} invocation, especially in complex apps with numerous dataflows, would inevitably induce permission fatigue as it already does today~\cite{choi2018role, pielot2018dismissed, bravo2014harder}. While a more comprehensive mechanism is needed to elicit dataflow preferences without overwhelming individuals, designing and rigorously evaluating such an interface is fundamentally a human-computer interaction (HCI) challenge. Consequently, solving this usability problem remains an important area for future work, one that relies on the technical foundation established by our escrow architecture.}

\section{Escrow Implementation} \label{sec:implementation}

This section details the technical implementation of the escrow's two core subcomponents: the relational engine that powers the on-device data virtualization layer, and the secure offloading pipeline that enables delegated computation on escrow-controlled servers. We describe the detailed techniques we employ to implement the escrow's relational engine using Virtual Tables with Pushdown (Section~\ref{sec:rel_engine_internals}) and its computation offloading pipeline (Section~\ref{sec:offloading_pipeline}).



\subsection{Relational Engine Internals} \label{sec:rel_engine_internals}

The core of our prototype is a relational engine that bridges the declarative world of SQL with the imperative, object-oriented APIs of Apple's native SDKs. We provide our implementation details of the Virtual Tables with Pushdown approach, as described in Section~\ref{sec:relational_stratedgies}. The virtual table implementation for each virtualized data source follows a consistent pattern: a thin C layer interfaces directly with SQLite, while a Swift backend contains the primary logic for data fetching and pushdown optimization.

\subsubsection{The C-to-Swift Virtual Table Bridge} We leverage SQLite's virtual table module~\cite{sqlite-vtab} to build the relational engine. For each data source (\eg Contacts, Photos, Location), we implement a standard \texttt{sqlite3\_module}~\cite{sqlite3_module}, a C structure that registers a set of callbacks with the SQLite engine. These C callbacks act as a lightweight bridge, delegating the main work to Swift functions exposed to the C runtime via the \texttt{@\_cdecl} attribute. This hybrid architecture allows us to combine the low-level control of SQLite's C API with the high-level power of Swift for interacting with modern Apple SDKs. The query lifecycle is managed by these key C callbacks:

\begin{myitemize}
    \item \texttt{xConnect}: Invoked by SQLite when creating the virtual table. We use them to declare the virtual table's schema (\eg the columns \texttt{givenName}, \texttt{familyName}, \texttt{phoneNumber} for the \texttt{Contacts} table).

    \item \texttt{xBestIndex}: This callback is the heart of our optimization logic. It is called by the SQLite query planner, which passes a representation of the query's constraints (\ie the \texttt{sqlite3\_index\_info} structure~\cite{sqlite3_index_info}), including information such as WHERE and LIKE clause predicates, ORDER BY terms, and LIMIT counts. Our \texttt{xBestIndex} implementation parses these constraints to determine which operations can be pushed down to native Apple SDKs, and encodes the optimized indexing strategy into an integer bitmask (for predicates) and a formatted string (for projection mask and other parameters) by filling in the \texttt{idxNum} and \texttt{idxStr} fields of the \texttt{sqlite3\_index\_info} structure, which forms the reply to the SQLite query planner.

    \item \texttt{xFilter}: This function executes the data fetch according to the indexing strategy selected in \texttt{xBestIndex}. Its primary role is to bridge into our Swift implementation to perform the data fetch, passing the predicate values and other details of the indexing strategy; the Swift logic is responsible for translating the encoded indexing strategy into an efficient data fetch request using native Apple SDKs. \texttt{xFilter} receives back an opaque pointer to a Swift object that contains an in-memory snapshot of the query results.
    
    \item \texttt{xNext}/\texttt{xColumn}: After \texttt{xFilter} fetches the result snapshot, SQLite uses this pair of functions to iterate through the result set. \texttt{xNext} simply increments a row counter, while \texttt{xColumn} calls back into our Swift implementation to retrieve the data for a specific row and column from the snapshot.    
\end{myitemize}


\subsubsection{Zero-Copy Native Object Handling} A key challenge is handling complex native Swift objects (\eg \texttt{PHAsset} or \texttt{CLLocation}). A naive approach, such as the one used in our Materialized Tables baseline, would require storing object identifiers in the database and performing costly lookups to reconstruct the native objects post-query. Our virtual table implementation avoids this overhead. When the \texttt{xColumn} callback iterates through query results, it obtains an opaque pointer to the Swift object in memory (via \texttt{Unmanaged.passRetained(object).toOpaque()}). This pointer is passed across the C bridge and what SQLite stores for the object-typed column. Later, when the escrow processes the query results from SQLite, this pointer is passed back from the C runtime to Swift and safely converted back into a fully-typed Swift object (via \texttt{Unmanaged.fromOpaque(pointer).takeRetainedValue()}). This zero-copy mechanism allows passing native objects through the relational engine without any serialization or expensive re-fetching, which is critical for the performance gains shown in our evaluation.

\subsection{Computation Offloading Pipeline} \label{sec:offloading_pipeline}

The escrow architecture supports offloading computation to escrow-controlled servers. This is enabled by an optional \textsc{server\_id} parameter in the \textsc{run} function call.

\subsubsection{Architecture and Security Model} The offloading pipeline runs on a client-server architecture where the on-device escrow acts as the client and the server is designed to run in a Trusted Execution Environment (TEE)~\cite{sabt2015trusted}, analogous to systems like Apple's Private Cloud Compute~\cite{private-cloud-compute}. Communication between client and server is encrypted using mutual TLS (mTLS), which authenticates both the device and server to protect data in transit. Furthermore, code integrity is enforced by requiring developers to pre-register and cryptographically sign any \textsc{compute()} function intended for offloading; the server verifies this signature against its registry before execution to prevent unauthorized code from running. These measures provide the technical foundation for securely extending the individual's trust zone to include remote infrastructure.

\subsubsection{End-to-End Offloading Pipeline} When an app calls \textsc{run(access(), compute(), \textsc{server\_id})}, it executes the following steps:

\mypar{1. Local Data Access} The \textsc{access()} always executes locally on the device via the relational engine to obtain the source data, $D_s$. This ensures raw data access remains under the local escrow's control.

\mypar{2. Serialization} The resulting Swift objects are serialized into a portable binary format (\eg Protocol Buffers~\cite{protobuf}).

\mypar{3. Secure Invocation} The on-device escrow establishes an mTLS connection to the server specified by \textsc{server\_id}. It then transmits a request payload containing the serialized $D_s$ and a unique identifier for the pre-registered \textsc{compute()} function.

\mypar{4. Remote Execution} The server authenticates the device's TLS certificate and verifies the integrity of the requested function. Upon success, it deserializes $D_s$ and invokes the corresponding \textsc{compute()} function within a sandboxed process, passing the data as input.

\mypar{5. Result Return} The final result, $D_t$, is serialized by the server and sent back to the device over the secure mTLS channel. The on-device escrow deserializes the result and returns it to the app.

This pipeline provides the technical foundation for enforcing resource-aware dataflow preferences, and enables the escrow to act as a scheduler, dynamically placing computation based on individuals' preferences, device state (\eg battery, network), and the computational cost of the dataflow.
\section{Evaluation}
\label{sec:evaluation}

In this section, we answer three research questions:

\begin{myitemize}
    \item \textbf{RQ1}: Can real-world applications be ported to the escrow? (\ie is the proposed escrow solution feasible?) We collect 10 representative apps and study their implementation on the escrow-based programming interface (Section~\ref{sec:rq1}).
    \item \textbf{RQ2:} Are escrow-based applications efficient? We measure the overhead the escrow introduces compared to applications written without using the escrow (Section~\ref{sec:rq2}).
    \item \textbf{RQ3:} What are the performance trade-offs of different implementations of the escrow's relational engine? We evaluate the performance of three approaches: Materialized Tables, Virtual Tables, and Virtual Tables with Pushdown (Section~\ref{sec:rq3}).
\end{myitemize}

\begin{table*}[]
\resizebox{\textwidth}{!}{%
\begin{tabular}{|c|c|c|c|c|c|c|c|c|c|c|}
\hline
\textbf{} &
  \textbf{Bluesky}\cite{Bluesky} &
  \textbf{Covid Alert}\cite{covid-alert} &
  \textbf{DuckDuckGo}\cite{DuckDuckGo} &
  \textbf{Nextcloud}\cite{Nextcloud} &
  \textbf{Signal}\cite{Signal} &
  \textbf{Simplenote}\cite{Simplenote} &
  \textbf{VLC}\cite{VLC} &
  \textbf{Wikipedia}\cite{Wikipedia} &
  \textbf{WordPress}\cite{WordPress} &
  \textbf{Home Assistant}\cite{Home-Assistant} \\ \hline
Category &
  Social media &
  Contact tracing &
  Browser &
  File Management &
  Messenger &
  Notes &
  Media player &
  Wikipedia &
  Blogs &
  Home automation \\ \hline
\begin{tabular}[c]{@{}c@{}}Contact \\ dataflow\end{tabular} &
  N/A &
  N/A &
  N/A &
  N/A &
  \begin{tabular}[c]{@{}c@{}}Find other users \\ from contacts\end{tabular} &
  \begin{tabular}[c]{@{}c@{}}Share notes with \\ other contacts\end{tabular} &
  N/A &
  N/A &
  N/A &
  N/A \\ \hline
\begin{tabular}[c]{@{}c@{}}Location\\ dataflow\end{tabular} &
  N/A &
  N/A &
  \begin{tabular}[c]{@{}c@{}}Share location\\ with websites to\\ obtain location\\ based results\end{tabular} &
  \begin{tabular}[c]{@{}c@{}}Show location \\ on a map\end{tabular} &
  \begin{tabular}[c]{@{}c@{}}Reverse geocode,\\ search and share \\ location\end{tabular} &
  N/A &
  N/A &
  \begin{tabular}[c]{@{}c@{}}Search and \\ recommend \\ articles based \\ on location\end{tabular} &
  \begin{tabular}[c]{@{}c@{}}Add current \\ location \\ to posts\end{tabular} &
  \begin{tabular}[c]{@{}c@{}}Send location \\ to local Home \\ Assistant \\ server\end{tabular} \\ \hline
\begin{tabular}[c]{@{}c@{}}Image/\\ video\\ dataflow\end{tabular} &
  \begin{tabular}[c]{@{}c@{}}Set profile \\ pictures, \\ create posts\end{tabular} &
  N/A &
  \begin{tabular}[c]{@{}c@{}}Perform voice \\ search\end{tabular} &
  \begin{tabular}[c]{@{}c@{}}Upload image/\\ video to \\ user’s cloud\end{tabular} &
  \begin{tabular}[c]{@{}c@{}}Send image/\\ video\end{tabular} &
  N/A &
  \begin{tabular}[c]{@{}c@{}}Control Apple \\ system event to \\ resume/pause media\end{tabular} &
  N/A &
  \begin{tabular}[c]{@{}c@{}}Add \\ photo/media \\ to posts\end{tabular} &
  \begin{tabular}[c]{@{}c@{}}Perform voice \\ assist\end{tabular} \\ \hline
\begin{tabular}[c]{@{}c@{}}Other\\ dataflow\end{tabular} &
  N/A &
  \begin{tabular}[c]{@{}c@{}}Record bluetooth\\ keys and match \\ keys with positive\\ Covid diagnosis\end{tabular} &
  N/A &
  N/A &
  \begin{tabular}[c]{@{}c@{}}Transfer app data\\  to another device \\ via LAN\end{tabular} &
  N/A &
  \begin{tabular}[c]{@{}c@{}}Access Desktop/\\ Documents/Downloads \\ folder, and network/\\ removable volume\end{tabular} &
  N/A &
  N/A &
  \begin{tabular}[c]{@{}c@{}}Sends motion \\ data and focus \\ status to\\ Home Assistant\end{tabular} \\ \hline
\end{tabular}%
}
\caption{Dataflows of representative apps. N/A means no dataflow exists in the app}
\label{tab:rq3}
\end{table*}

\subsection{RQ1: Dataflows in Real-World Apps} \label{sec:rq1}

We design a rubric to collect a representative sample of iOS apps containing various dataflows involving sensitive personal data, and we describe how those dataflows can be reimplemented using the escrow's programming interface.
    
\mypar{Rubrics for selecting apps} We selected 10 representative apps based on the following rubric~\footnote{We include Covid Alert (Canada's COVID contact tracing app)~\cite{covid-alert}, which does not fit all rubrics but has interesting dataflows that are worth discussing in the paper.}:

\begin{myitemize}
    \item Each app needs to be open-sourced in GitHub so we can find dataflows in its codebase, and it should be currently maintained (\ie not archived and has commits in the recent month) 
    \item Each app needs to be relatively popular for it to be representative, thus we only select apps with over 1000 Github stars.
    \item Apps need to contain dataflows covering different types of data.
    \item Apps need to be diverse, spanning multiple categories.
\end{myitemize} 

\mypar{Analysis} We analyze each app's source code, identify dataflows involving sensitive personal data, and summarize our findings in Table~\ref{tab:rq3}. We include one row describing the dataflow involving contacts, location, and image/video respectively, since most apps contain dataflows that involve at least one of the three data types. We also include one row describing other dataflows.

\mypar{Location dataflow} Six apps use individuals' current location for various purposes. Five of them (DuckDuckGo~\cite{DuckDuckGo}, Signal~\cite{Signal}, Wikipedia~\cite{Wikipedia}, Home Assistant~\cite{Home-Assistant}, WordPress~\cite{WordPress}) share the location outside the individuals' devices with servers. For instance, Wikipedia finds and recommends articles that individuals might be interested in based on their surrounding region. The app makes a search request to Wikimedia (the parent company of Wikipedia) server with the individual's surrounding region as input, and the server returns relevant articles. We show a simplified version of how the dataflow can be implemented using the escrow based on Wikipedia's current implementation:

\begin{minted}{swift}
Escrow.run("SELECT longitude, latitude FROM Location
ORDER BY timestamp LIMIT 1") { 
coordinates -> SearchResults in
    let region = getSurroundingRegion(
        location: coordinates, radius: 1000)
    // Create search request using Wikimedia's API
    let url = wikiMediaAPI(region: region)
    let results = fetchArticles(url: url, limit: 50)
    return results }
\end{minted}

Although the exact details of how Wikimedia finds articles based on a given location are unknown, implementing the dataflow using the escrow makes it transparent exactly \emph{what} data is leaving the individual's device and its purpose through \textsc{compute()}. In the case of Wikipedia, the search request only needs the approximate region that centers around the individual's current location by a specified radius, and the precise location coordinates or other identifying information should not leave the device. The app only has access to the fetched articles as the output of \textsc{run} and nothing else.

\mypar{Contact dataflow} Two apps, Signal~\cite{Signal} and Simplenote~\cite{Simplenote}, contain dataflows involving contacts. Signal performs "contact discovery" by fetching the individual's contacts (specifically phone numbers) and finding which contacts are Signal users. Notably, the computation takes place on Signal servers, which run in a secure hardware enclave with oblivious RAM to ensure Signal or anyone else does not gain access to individuals' contact data~\cite{signal-enclave}; essentially, the computation is equivalent to a private set intersection between individuals' contacts and Signal's users~\cite{popa2024confidential}. A simplified implementation of this dataflow using the escrow may look like: 

\begin{minted}{swift}
Escrow.run("SELECT phoneNumber FROM Contacts") { 
phoneNumbers -> SignalAccounts in
    let url = SignalEnclaveURL()
    let matched = contactDiscovery(phoneNumbers, url)
    return matched }
\end{minted}

While technically the mechanisms Signal implemented for contact discovery would already ensure individuals' contact data remain private and not accessible to anyone else, implementing this dataflow via the escrow makes it transparent that this computation \emph{only} takes place in Signal's enclave and the data is not used for other purposes. In addition, it is conceivable that the contact discovery can take place in an escrow-controlled server with similar enclave technologies, so individuals gain more control over the specific usage of their data. Lastly, it is worth noting that such emphasis on privacy-preserving computation, like what Signal has implemented, is \emph{not} the current industry norm~\cite{zuo2019does}. 



On the other hand, Simplenote lets individuals pick which contact they would like to share their notes with, then sends a request to the contact (\eg via email). This dataflow can also be implemented using the escrow; for instance, \textsc{access()} fetches necessary contact records (\eg name and email), and \textsc{compute()} implements the interface that allows the individual to pick which contact to share with, then sends the request to the contact. The escrow makes it transparent what data Simplenote has access to (in this case it would only have access to the specific contact the individual picked) and how the data is used.

\begin{table*}[ht!]
\centering
\resizebox{\textwidth}{!}{%
\begin{tabular}{|c|c|c|}
\hline
 &
  Data Access &
  Computation \\ \hline
Contacts &
  \begin{tabular}[c]{@{}c@{}}Get the phone number of the contact whose given name is 'uniqueName'\\ \texttt{SELECT phoneNumber FROM Contacts WHERE givenName = 'uniqueName'}\end{tabular} &
  Check if the result is a valid US phone number \\ \hline
Photos &
  \begin{tabular}[c]{@{}c@{}}Get the most recent 100 images (as \texttt{PHAsset}s) from the Photos library\\ \texttt{SELECT phasset FROM Photos WHERE mediaType=='image' ORDER BY creationDate DESC LIMIT 100}\end{tabular} &
  Transform the result to a list of \texttt{NSImage} \\ \hline
Location &
  \begin{tabular}[c]{@{}c@{}}Get the most recent location (as \texttt{CLLocation})\\ \texttt{SELECT cllocation from Location ORDER BY timestamp DESC LIMIT 1}\end{tabular} &
  Get weather using OpenWeatherMap~\cite{openweathermap} API \\ \hline
\end{tabular}%
}
\caption{Dataflows run by the escrow-based and baseline app, which uses SQL query and Apple SDKs to access data respectively}
\label{tab:rq2_dataflow}
\end{table*}

\begin{figure*}[ht!]
    \centering
    \includegraphics[width=\textwidth]{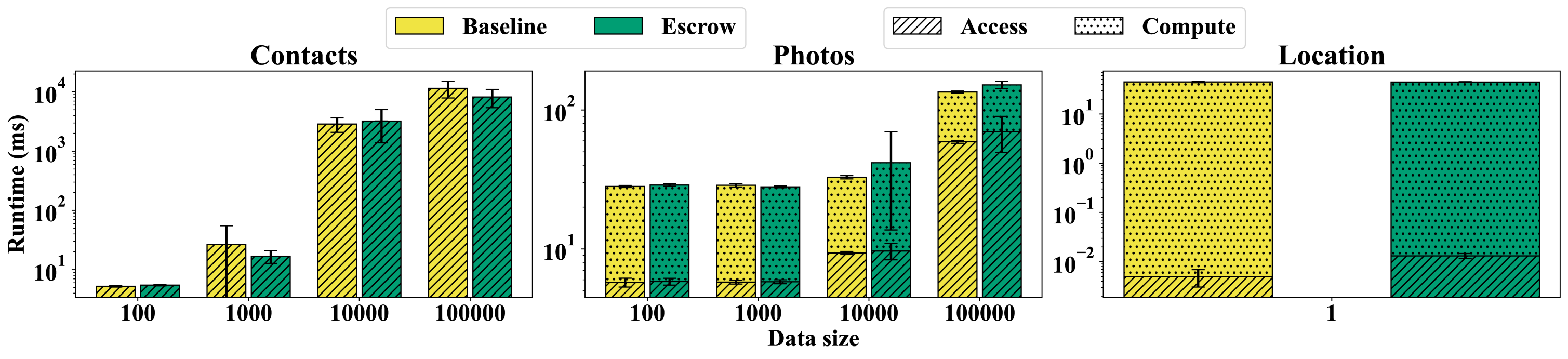}
    \caption{Average access and compute runtime (with std) by escrow-based app vs baseline app over 10 runs}
    \label{fig:rq2}
\end{figure*}

\mypar{Image/video dataflow} Seven apps have read access to the device's Photos Library. Most apps access the Photo Library for functionalities such as setting avatars (Bluesky~\cite{Bluesky}), creating posts (Bluesky, WordPress~\cite{WordPress}), sending / uploading messages / files (Nextcloud~\cite{Nextcloud}, Signal), or playing videos (VLC~\cite{VLC}). We concentrate on one unique use case from two apps (DuckDuckGo, Home Assistant), which is speech recognition. Generally, there are two ways to perform speech recognition: one is to transcribe a recording that's already stored on-device, and the other is to transcribe live speech. Implementing the former task using the escrow is straightforward; for example, to transcribe the most recently stored audio recording, developers can write:

\begin{minted}{swift}
Escrow.run("SELECT asset FROM Photos WHERE mediaType
=='audio' ORDER BY creationDate LIMIT 1") { 
recording -> String in
    let recognizer = SpeechRecognizer()
    let text = recognizer.transcribe(recording)
    return text }
\end{minted}

The SQL query fetches the most recently created asset stored in the Photos library that has an audio media type. The query returns a native \texttt{PHAsset} object that can be directly consumed by the \texttt{SpeechRecognizer}. On the other hand, transcribing a live speech is generally achieved by storing the individual's live speech in a buffer in memory. When the buffer is full (or after a small time interval), the speech recognizer will transcribe the new buffer (along with the previously stored buffers), thus achieving the effect of transcribing speech on the fly. Since no speech data is physically stored on-device, the escrow cannot access the data, thus we consider implementing this dataflow to be outside the scope of this paper. However, it is possible that by integrating with Apple's infrastructure the escrow will be able to access structured data that is stored in memory and enable dataflows such as live speech transcription. 

\mypar{Other dataflows} Four apps contain dataflows that do not involve contact, location, or image/video data, each with its distinct purpose. All dataflows from Covid Alert, VLC, and Home Assistant can be implemented using the escrow except Signal, since Signal's dataflow involves app-specific data that is not accessible using Apple SDKs. We explain how to implement the dataflow from Covid Alert, which we consider to be the most interesting one among the four apps. 

Covid Alert is Canada's official Covid contact tracing app. The app uses Apple's Exposure Notification framework~\cite{contacttracing}. It works by recording each individual's on-device Bluetooth keys when their device exchanges Bluetooth beacons with other devices. When an individual receives a positive COVID diagnosis, they can voluntarily share their keys with the government-controlled remote key server, which stores all Bluetooth keys from individuals who have recently tested positive for COVID. Periodically, the app downloads keys from the key server and tries to find whether the individual's keys from the last 14 days match any of the downloaded keys (from those who tested positive). If there is a match, the app notifies the individual. The two important dataflows---sharing keys with the key server when an individual tests positive, and matching the individual's keys with keys downloaded from the key server, can both be implemented using the escrow. For example, an approximate implementation of the former dataflow may look like:

\begin{minted}{swift}
Escrow.run("SELECT * FROM KeyTable ORDER BY data LIMIT 
14") { keys in let url = keyServerURL()
    shareKeysWithServer(url) }
\end{minted}

Using the escrow provides transparency of the dataflow and ensures that the individual's private health information is not exposed to anyone else other than the government-controlled key server.

In summary, we show that it is feasible to re-implement many dataflows in real-world apps via the escrow's programming interface. We note that since the apps we collected are all open-source, they are a "biased" sample in the sense that the apps are already designed with transparency as a priority. Many closed-source apps have undisclosed dataflows for purposes not covered in our analysis (\eg profiling, resale). Enforcing those dataflows to be implemented using the escrow would enable individuals to gain transparency and control over these dataflows.   

\subsection{RQ2: Runtime Efficiency of the Escrow} 
\label{sec:rq2}

\begin{table*}[ht!]
\centering
\resizebox{\textwidth}{!}{%
\begin{tabular}{|c|c|c|c|c|}
\hline
         & Full             & Projection       & Predicate        & Order By / Limit                                                                                  \\ \hline
Contacts &
  \texttt{SELECT * FROM Contacts} &
  \begin{tabular}[c]{@{}c@{}}\texttt{SELECT familyName, givenName}\\\texttt{FROM Contacts}\end{tabular} &
  \begin{tabular}[c]{@{}c@{}}\texttt{SELECT * FROM Contacts}\\\texttt{WHERE givenName = 'uniqueName'}\end{tabular} &
  \textbackslash{} \\ \hline
Photos &
  \texttt{SELECT * FROM Photos} &
  \texttt{SELECT phasset FROM Photos} &
  \begin{tabular}[c]{@{}c@{}}\texttt{SELECT * FROM Photos}\\\texttt{WHERE collectionName = 'uniqueAlbum'}\end{tabular} &
  \begin{tabular}[c]{@{}c@{}}\texttt{SELECT * FROM Photos ORDER BY}\\\texttt{creationDate DESC LIMIT 1}\end{tabular} \\ \hline
Location & \textbackslash{} & \textbackslash{} & \textbackslash{} & \begin{tabular}[c]{@{}c@{}}\texttt{SELECT * FROM Location ORDER BY}\\\texttt{timestamp DESC LIMIT 1}\end{tabular} \\ \hline
\end{tabular}%
}
\caption{Queries run by the escrow's relational interface}
\label{tab:rq3_queries}
\end{table*}

\begin{figure*}[ht!]
    \centering
    \includegraphics[width=\textwidth]{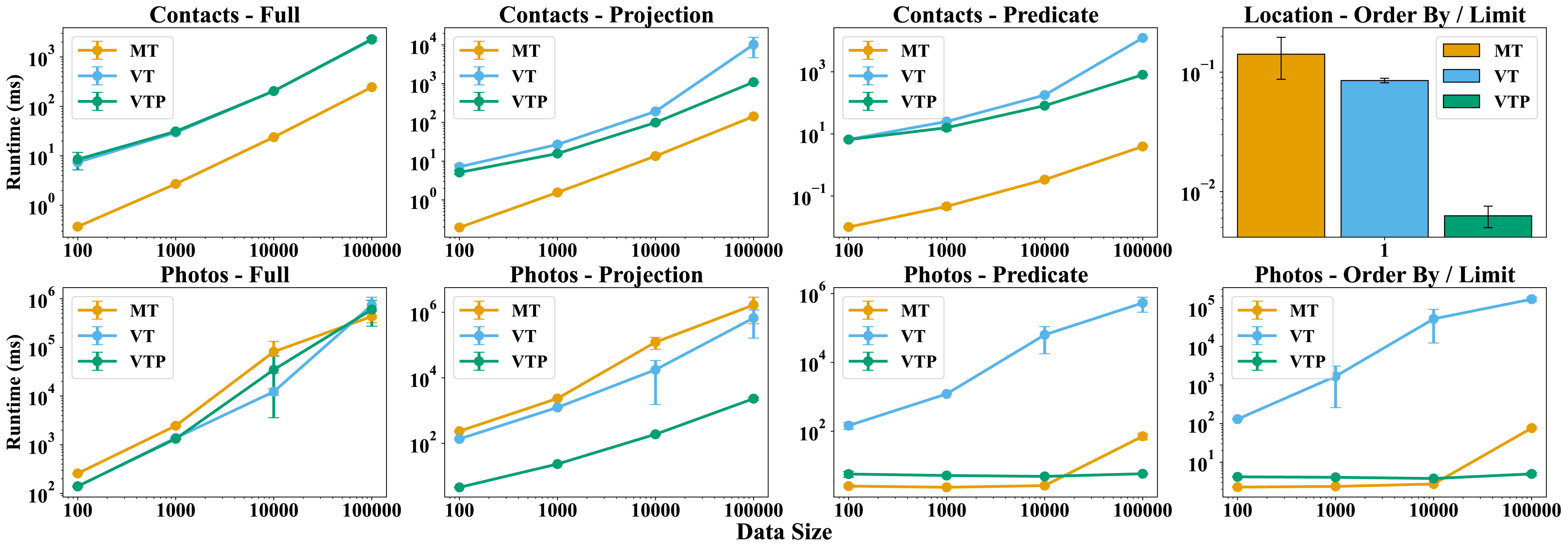}
    \caption{Average runtime (with std) to execute queries over 10 runs using Materialized Tables (MT), Virtual Tables (VT), and Virtual Tables with Pushdown (VTP)}
    \label{fig:rel_bench}
\end{figure*}

We measure the overhead introduced by the escrow by comparing the time to access data and run computation on the data using the escrow's \textsc{run} function versus without the escrow (\ie the baseline) for different data sizes. Table~\ref{tab:rq2_dataflow} shows the dataflows used in our evaluation, which involve three data types that are representative in many of today's apps, as we have shown in Section~\ref{sec:rq1}.

\mypar{Implementation} We implement an escrow-based app, which uses the escrow's \textsc{run} function to access data and run computation, and the baseline app, which uses Apple SDKs to access data. The computation functions in both apps are implemented in a similar way and the output of the computation are exactly the same. 

The escrow-based app uses the escrow as a static singleton instance (note that the escrow prototype is deployed as a library). We implement its relational engine using Virtual Tables with Pushdown. The schemas of the virtual tables are:

\begin{myitemize}
    \small
    \item \texttt{Contacts}: \texttt{id TEXT, givenName TEXT, familyName TEXT, phoneNumber TEXT}

    \item \texttt{Photos}: \texttt{id TEXT, phasset PHAsset, mediaType INT, creationDate DATE, collectionID TEXT, collectionName TEXT} 

    \item \texttt{Location}: \texttt{cllocation CLLocation, longitude REAL, latitude\\REAL, horizontalAccuracy REAL, timestamp TIMESTAMP}
\end{myitemize}

We implement projection pushdowns for all three virtual tables. In addition, for Contacts table, we implement predicate pushdown for all four columns, For Photos table, we implement predicate pushdown for \texttt{id}, \texttt{mediaType}, \texttt{collectionID} and \texttt{collectionName}, and order by / limit pushdown for \texttt{creationDate}. For Location table, we implement order by / limit pushdown for \texttt{timestamp}.

\mypar{Data preparation} We prepare and run the experiments for different data sizes. For contacts, we add 100, 1k, 10k, and 100k contacts records to the Contacts app; each record has a unique given name, family name, and phone number. For photos, we add 100, 1k, 10k, and 100k photos to the Photos library; each photo is 16KB. Location data is updated in real time.

\mypar{Configurations} The apps are implemented as multi-platform (iOS/macOS) apps in Swift 6.1. We deploy the app as a macOS app and run the experiments using a 2021 MacBook Pro with Apple M1 Pro chip, 16 GB memory, and macOS 15.6.1, since iPhones cannot handle the large data volumes we will insert.

\mypar{Result} We run each app 10 times for each dataflow and data size, and we show the average access and computation time separately (with standard deviation) in Figure~\ref{fig:rq2}. In short, the escrow introduces very little overhead. For contacts dataflows, the runtime is dominated by the data access time, since checking whether a given number is a valid US phone number only involves regex matching and is very fast. Because the escrow executes the SQL query via the Contacts virtual table, which implements pushdown for projections and predicates, the database does not perform additional data filtering, hence the overhead is small compared to fetching data directly using Apple's Contacts SDK. For photos dataflow, as data size increases, the majority of runtime shifts from computation to data access, since the number of photos to process stays the same in the computation function. Similarly to Contacts, the Photos virtual table also implements pushdown so there is no additional data filtering on the database side. For Location dataflows, the runtime is dominated by the computation time, since data is transferred to and from the OpenWeatherMap server. 

In summary, the escrow's overhead comes from the relational engine and any post-processing the escrow needs to perform to transform the query result to a unified format so it can be processed by the subsequent computation function. Implementing pushdown in virtual tables reduces the amount of data filtering the database needs to do, and implementing the zero-copy mechanism for passing native Swift objects through the relational engine ensures there is no serialization overhead, thus reducing the overall overhead. 

\subsection{RQ3: Relational Engine Comparison} \label{sec:rq3}

We compare the runtime performance of three implementations of the escrow's relational engine (Section~\ref{sec:relational}): Materialized Tables, Virtual Tables, and Virtual Tables with Pushdown. Table~\ref{tab:rq3_queries} shows the queries used in the evaluation. We categorize the queries into full, projection, predicate and order by / limit queries in order to evaluate how each implementation performs for each category. 

\mypar{Setup} All three implementations are exposed to the same schema as described in the previous experiment (Section~\ref{sec:rq2}). We also use the same data preparation and experiment configurations as in the previous experiment. We create a unique album with one photo inside, in addition to the photos we have already added. We only measure the performance of running the \emph{data access} function and there is no computation function involved. All three implementations produce query result in the same format.

\begin{table}[h]
    \centering
    \begin{tabular}{|c|c|c|c|c|}
    \hline
         & size=100 & size=1000 & size=10000 & size=100000\\
         \hline
        Contacts & 10.85 & 32.68 & 279.28 & 3402.41\\
        \hline
        Photos & 290.72 & 1714.91 & 19150.86 & 130038.08\\
        \hline
    \end{tabular}
    \caption{Time (ms) to materialize Contacts and Photos table}
    \label{tab:materialize_time}
    \vspace{-20pt}
\end{table}

\mypar{Result} We run each query 10 times, and we show the average runtime with standard deviation in Figure~\ref{fig:rel_bench}. In addition, for Materialized Tables, we report the time to materialize the Contacts and Photos table for each data size in Table~\ref{tab:materialize_time}. The time grows linearly with the data size. Materializing the Photos table takes considerably more time since it needs to enumerate each photo asset and fetch its collection ID and name. The runtime performance of the three implementations varies for different categories of queries. As expected, Virtual Tables with Pushdown perform better than Virtual Tables for all except full (SELECT *) queries, in which case the two implementations have roughly the same runtime since no pushdown can be applied. Notably, Virtual Tables with Pushdown performs magnitudes faster for highly selective queries (Photos queries with predicate and order / limit), because it asks the \texttt{PhotoKit} framework for a single album with only one photo inside or the single most recent photo, avoiding the materialization of unnecessary data, whereas Virtual Tables implementation must filter or sort all photos. It is also the fastest for Photos queries with projection since the other two implementations both need to fetch collection information for each \texttt{phasset}, which is costly. Materialized Tables implementation is about an order of magnitude faster than the other two for Contacts queries but not for Photos queries, because for Contacts queries the data is already in the database and SQLite can employ any internal optimization it needs to execute the query efficiently, but for Photos queries the escrow needs to map the result \texttt{phasset} IDs to \texttt{phasset} objects when the query finalizes, which introduces overhead. Both Virtual Tables implementations use zero-copy mechanism to pass \texttt{phasset} objects through the database thus do not incur serialization overhead.

In summary, while Materialized Tables offers speed for full scans, it suffers from high initialization costs and data staleness. By applying classic query optimization principles, Virtual Tables with Pushdown provides the data freshness of on-the-fly data virtualization with performance that is competitive with, and in some cases superior to, querying materialized data.

\section{Related Work} 
\label{sec:relatedworks}

\mypar{Personal Data Sovereignty} The concept of personal data sovereignty has emerged as a critical paradigm for rebalancing control in the digital economy~\cite{ernstberger2023sok, polvcak2017information, hummel2021data, lauf2022linking}. 
Our work's central goal of \emph{Personal Dataflow Sovereignty} is a direct, technical instantiation of this principle, providing a concrete mechanism for individuals to exercise control over how their data is used. While legal frameworks like the GDPR~\cite{gdpr} provide a foundation for personal data sovereignty, their practical implementation often falls short~\cite{ryan2024will}, highlighting the need for a technical foundation like the one we propose. The data management community has a rich history of engaging with the core technical problems under personal data sovereignty, from \emph{Hippocratic Databases}~\cite{agrawal2002hippocratic} that articulates a vision for database systems designed with privacy as a core principle, to seminal work on purpose-based access control databases~\cite{byun2008purpose}, to more recently a data escrow system~\cite{xia2023data} that enables delegated computation.

\mypar{Data Governance Frameworks} The concept of rebalancing power in the data economy has given rise to various data governance frameworks~\cite{abraham2019data}. These include Data Unions, Data Trusts, and Data Cooperatives, which advocate for collective organizations that would manage data and negotiate on behalf of individuals~\cite{delacroix2019bottom, mozilla-shifting-power, ada2021exploring}. A related economic perspective reframes individual-generated data as a form of digital labor, arguing that individuals should be compensated for their contributions to the data economy, a concept known as Data as Labor~\cite{posner2018radical, arrieta2018should}. While these frameworks provide compelling visions for a more equitable data ecosystem, they often lack a concrete technical foundation for their implementation. Our data escrow architecture can be viewed as a critical piece of enabling the technical infrastructure that serves as the trustworthy intermediary required for such frameworks to operate effectively.

\mypar{Privacy design and enforcement mechanism} A large body of research focuses on usable privacy and policy design in apps. 
For instance, many works focus on privacy designs to guide individuals toward informed privacy choices~\cite{acquisti2017nudges, zhang2023privacy, feng2021design, zhang2022usable, habib2022evaluating, schaub2015design, zhang2024exploring}. Recent research has also explored the use of Large Language Models (LLMs) to facilitate privacy comprehension~\cite{duesterwald2025can, feng2024understanding}. However, a critical gap exists: usable privacy solutions frequently lack the underlying technical feasibility to be deployed effectively in practice~\cite{iravantchi2025sok, brodie2005usable, fischer2024overview, distler2021systematic, pan2024new}. Our escrow provides the technical \emph{mechanism} required to enforce individuals' dataflow preferences, thereby motivating and providing the necessary foundation for future \emph{policy} design on eliciting those preferences. \update[MR6]{In addition, existing industry practices offer alternative approaches to data minimization, such as data pickers that allow granting apps access to only the subset of data individuals have picked (\eg Apple's Photo picker~\cite{phpicker}). However, they fundamentally constrain the types of computations a developer can implement. The escrow model differs by allowing developers to specify arbitrary computation over data.}

\mypar{Relational model over heterogeneous data sources} The challenge of providing a unified query interface over multiple heterogeneous data sources is a classic problem in the database community~\cite{halevy2001answering}.  Data virtualization techniques are pioneered by early data integration systems like Garlic~\cite{carey1995towards}, where source-specific wrappers translate queries against a global schema into the native languages or API calls of the underlying sources. This principle is now widely adopted in production systems, for instance, through PostgreSQL's Foreign Data Wrapper (FDW)~\cite{postgres-fdw}. The escrow's relational interface applies this same principle of data virtualization, creating a unified schema over disparate, programmatic APIs to access data. Our approach is also analogous to how TinyDB~\cite{madden2005tinydb} overlays a relational model over data collected from sensor networks.

\mypar{\update{Trusted Execution and Computation Offloading}} \update[MR7]{A core capability of the escrow is offloading computation while keeping data within the individual's trust zone, necessitating mechanisms for secure execution. Therefore, we contextualize our architecture alongside prior systems that secure data and computation against untrusted environments. In cloud settings, systems like Sieve~\cite{wang2016sieve} and Iron~\cite{fisch2017iron} leverage functional encryption and cryptographic access controls, while Slalom~\cite{tramer2018slalom} and Telekine~\cite{hunt2020telekine} use hardware enclaves to secure remote workloads. On mobile devices, frameworks such as TinMan~\cite{xia2015tinman}, ShadowNet~\cite{sun2023shadownet}, and SeCloak~\cite{lentz2018secloak} rely on hardware isolation to shield peripheral data and execution from the host OS. While the escrow architecture provides the orchestration layer, which dictates whether a dataflow is permitted based on its purpose, these prior systems provide the execution layer, which dictates how it is securely computed. When the escrow executes a \textsc{compute()} function on local hardware or offloads it to an escrow-controlled server, it could leverage these technologies to ensure the execution remains trustworthy.}

\section{Conclusion} \label{sec:conclusion}

In this paper, we contribute a solution for a critical problem in the modern data economy: the loss of personal data sovereignty. We propose a data escrow architecture that leverages delegated computation. We design and implement a minimally invasive programming interface and an efficient data virtualization layer to make this model practical and developer-friendly. Our concrete implementation in the Apple ecosystem demonstrates a feasible path to deployment. Finally, our comprehensive evaluation shows both qualitatively and quantitatively that our escrow solution is expressive enough to implement dataflows in a wide range of real-world applications, and that it is efficient.

\begin{acks}
We thank the anonymous reviewers for their detailed and constructive feedback, and the Data Ecology Research Initiative of the Data Science Institute for their support.
\end{acks}

\bibliographystyle{ACM-Reference-Format}
\bibliography{main}


\begin{thebibliography}{101}


\ifx \showCODEN    \undefined \def \showCODEN     #1{\unskip}     \fi
\ifx \showDOI      \undefined \def \showDOI       #1{#1}\fi
\ifx \showISBNx    \undefined \def \showISBNx     #1{\unskip}     \fi
\ifx \showISBNxiii \undefined \def \showISBNxiii  #1{\unskip}     \fi
\ifx \showISSN     \undefined \def \showISSN      #1{\unskip}     \fi
\ifx \showLCCN     \undefined \def \showLCCN      #1{\unskip}     \fi
\ifx \shownote     \undefined \def \shownote      #1{#1}          \fi
\ifx \showarticletitle \undefined \def \showarticletitle #1{#1}   \fi
\ifx \showURL      \undefined \def \showURL       {\relax}        \fi
\providecommand\bibfield[2]{#2}
\providecommand\bibinfo[2]{#2}
\providecommand\natexlab[1]{#1}
\providecommand\showeprint[2][]{arXiv:#2}

\bibitem[Abraham et~al\mbox{.}(2019)]%
        {abraham2019data}
\bibfield{author}{\bibinfo{person}{Rene Abraham}, \bibinfo{person}{Johannes Schneider}, {and} \bibinfo{person}{Jan Vom~Brocke}.} \bibinfo{year}{2019}\natexlab{}.
\newblock \showarticletitle{Data governance: A conceptual framework, structured review, and research agenda}.
\newblock \bibinfo{journal}{\emph{International journal of information management}}  \bibinfo{volume}{49} (\bibinfo{year}{2019}), \bibinfo{pages}{424--438}.
\newblock


\bibitem[Acar et~al\mbox{.}(2018)]%
        {acar2018survey}
\bibfield{author}{\bibinfo{person}{Abbas Acar}, \bibinfo{person}{Hidayet Aksu}, \bibinfo{person}{A~Selcuk Uluagac}, {and} \bibinfo{person}{Mauro Conti}.} \bibinfo{year}{2018}\natexlab{}.
\newblock \showarticletitle{A survey on homomorphic encryption schemes: Theory and implementation}.
\newblock \bibinfo{journal}{\emph{ACM Computing Surveys (Csur)}} \bibinfo{volume}{51}, \bibinfo{number}{4} (\bibinfo{year}{2018}), \bibinfo{pages}{1--35}.
\newblock


\bibitem[Acquisti et~al\mbox{.}(2017)]%
        {acquisti2017nudges}
\bibfield{author}{\bibinfo{person}{Alessandro Acquisti}, \bibinfo{person}{Idris Adjerid}, \bibinfo{person}{Rebecca Balebako}, \bibinfo{person}{Laura Brandimarte}, \bibinfo{person}{Lorrie~Faith Cranor}, \bibinfo{person}{Saranga Komanduri}, \bibinfo{person}{Pedro~Giovanni Leon}, \bibinfo{person}{Norman Sadeh}, \bibinfo{person}{Florian Schaub}, \bibinfo{person}{Manya Sleeper}, {et~al\mbox{.}}} \bibinfo{year}{2017}\natexlab{}.
\newblock \showarticletitle{Nudges for privacy and security: Understanding and assisting users’ choices online}.
\newblock \bibinfo{journal}{\emph{ACM Computing Surveys (CSUR)}} \bibinfo{volume}{50}, \bibinfo{number}{3} (\bibinfo{year}{2017}), \bibinfo{pages}{1--41}.
\newblock


\bibitem[Agrawal et~al\mbox{.}(2002)]%
        {agrawal2002hippocratic}
\bibfield{author}{\bibinfo{person}{Rakesh Agrawal}, \bibinfo{person}{Jerry Kiernan}, \bibinfo{person}{Ramakrishnan Srikant}, {and} \bibinfo{person}{Yirong Xu}.} \bibinfo{year}{2002}\natexlab{}.
\newblock \showarticletitle{Hippocratic databases}. In \bibinfo{booktitle}{\emph{VLDB'02: Proceedings of the 28th International Conference on Very Large Databases}}. Elsevier, \bibinfo{pages}{143--154}.
\newblock


\bibitem[Almuhimedi et~al\mbox{.}(2015)]%
        {almuhimedi2015your}
\bibfield{author}{\bibinfo{person}{Hazim Almuhimedi}, \bibinfo{person}{Florian Schaub}, \bibinfo{person}{Norman Sadeh}, \bibinfo{person}{Idris Adjerid}, \bibinfo{person}{Alessandro Acquisti}, \bibinfo{person}{Joshua Gluck}, \bibinfo{person}{Lorrie~Faith Cranor}, {and} \bibinfo{person}{Yuvraj Agarwal}.} \bibinfo{year}{2015}\natexlab{}.
\newblock \showarticletitle{Your location has been shared 5,398 times! A field study on mobile app privacy nudging}. In \bibinfo{booktitle}{\emph{Proceedings of the 33rd annual ACM conference on human factors in computing systems}}. \bibinfo{pages}{787--796}.
\newblock


\bibitem[Apple(2024)]%
        {private-cloud-compute}
\bibfield{author}{\bibinfo{person}{Apple}.} \bibinfo{year}{2024}\natexlab{}.
\newblock \bibinfo{booktitle}{\emph{Private Cloud Compute: A new frontier for AI privacy in the cloud}}.
\newblock
\urldef\tempurl%
\url{https://security.apple.com/blog/private-cloud-compute/}
\showURL{%
\tempurl}


\bibitem[Apple(2025a)]%
        {cllocation}
\bibfield{author}{\bibinfo{person}{Apple}.} \bibinfo{year}{2025}\natexlab{a}.
\newblock \bibinfo{booktitle}{\emph{CLLocation}}.
\newblock
\urldef\tempurl%
\url{https://developer.apple.com/documentation/corelocation/cllocation}
\showURL{%
\tempurl}


\bibitem[Apple(2025b)]%
        {cncontactfetchrequest}
\bibfield{author}{\bibinfo{person}{Apple}.} \bibinfo{year}{2025}\natexlab{b}.
\newblock \bibinfo{booktitle}{\emph{CNContactFetchRequest}}.
\newblock
\urldef\tempurl%
\url{https://developer.apple.com/documentation/contacts/cncontactfetchrequest}
\showURL{%
\tempurl}


\bibitem[Apple(2025c)]%
        {swift-contacts}
\bibfield{author}{\bibinfo{person}{Apple}.} \bibinfo{year}{2025}\natexlab{c}.
\newblock \bibinfo{booktitle}{\emph{Contacts Framework}}.
\newblock
\urldef\tempurl%
\url{https://developer.apple.com/documentation/contacts}
\showURL{%
\tempurl}


\bibitem[Apple(2025d)]%
        {phasset}
\bibfield{author}{\bibinfo{person}{Apple}.} \bibinfo{year}{2025}\natexlab{d}.
\newblock \bibinfo{booktitle}{\emph{PHAsset}}.
\newblock
\urldef\tempurl%
\url{https://developer.apple.com/documentation/photokit/phasset}
\showURL{%
\tempurl}


\bibitem[Apple(2025e)]%
        {photokit}
\bibfield{author}{\bibinfo{person}{Apple}.} \bibinfo{year}{2025}\natexlab{e}.
\newblock \bibinfo{booktitle}{\emph{PhotoKit}}.
\newblock
\urldef\tempurl%
\url{https://developer.apple.com/documentation/photokit}
\showURL{%
\tempurl}


\bibitem[Apple(2026a)]%
        {apple-capabilities}
\bibfield{author}{\bibinfo{person}{Apple}.} \bibinfo{year}{2026}\natexlab{a}.
\newblock \bibinfo{booktitle}{\emph{Adding capabilities to your app}}.
\newblock
\urldef\tempurl%
\url{https://developer.apple.com/documentation/xcode/adding-capabilities-to-your-app}
\showURL{%
\tempurl}


\bibitem[Apple(2026b)]%
        {app-review}
\bibfield{author}{\bibinfo{person}{Apple}.} \bibinfo{year}{2026}\natexlab{b}.
\newblock \bibinfo{booktitle}{\emph{App Review Guidelines}}.
\newblock
\urldef\tempurl%
\url{https://developer.apple.com/app-store/review/guidelines/}
\showURL{%
\tempurl}


\bibitem[Apple(2026c)]%
        {swift-core-location}
\bibfield{author}{\bibinfo{person}{Apple}.} \bibinfo{year}{2026}\natexlab{c}.
\newblock \bibinfo{booktitle}{\emph{Core Location Framework}}.
\newblock
\urldef\tempurl%
\url{https://developer.apple.com/documentation/corelocation}
\showURL{%
\tempurl}


\bibitem[Apple(2026d)]%
        {apple-entitlements}
\bibfield{author}{\bibinfo{person}{Apple}.} \bibinfo{year}{2026}\natexlab{d}.
\newblock \bibinfo{booktitle}{\emph{Entitlements}}.
\newblock
\urldef\tempurl%
\url{https://developer.apple.com/documentation/bundleresources/entitlements}
\showURL{%
\tempurl}


\bibitem[Apple(2026e)]%
        {apple-privacy-manifest}
\bibfield{author}{\bibinfo{person}{Apple}.} \bibinfo{year}{2026}\natexlab{e}.
\newblock \bibinfo{booktitle}{\emph{Privacy manifest files}}.
\newblock
\urldef\tempurl%
\url{https://developer.apple.com/documentation/bundleresources/privacy-manifest-files}
\showURL{%
\tempurl}


\bibitem[Apple(2026f)]%
        {apple-purpose-string}
\bibfield{author}{\bibinfo{person}{Apple}.} \bibinfo{year}{2026}\natexlab{f}.
\newblock \bibinfo{booktitle}{\emph{Requesting access to protected resources}}.
\newblock
\urldef\tempurl%
\url{https://developer.apple.com/documentation/uikit/requesting-access-to-protected-resources}
\showURL{%
\tempurl}


\bibitem[Apple(2026g)]%
        {phpicker}
\bibfield{author}{\bibinfo{person}{Apple}.} \bibinfo{year}{2026}\natexlab{g}.
\newblock \bibinfo{booktitle}{\emph{Selecting Photos and Videos in iOS}}.
\newblock
\urldef\tempurl%
\url{https://developer.apple.com/documentation/photokit/selecting-photos-and-videos-in-ios}
\showURL{%
\tempurl}


\bibitem[Apple(2026h)]%
        {apple-tcc}
\bibfield{author}{\bibinfo{person}{Apple}.} \bibinfo{year}{2026}\natexlab{h}.
\newblock \bibinfo{booktitle}{\emph{Transparency, Consent, and Control (TCC) Target Flag}}.
\newblock
\urldef\tempurl%
\url{https://security.apple.com/bounty/target-flags/#tcc}
\showURL{%
\tempurl}


\bibitem[Apple and Google(2025)]%
        {contacttracing}
\bibfield{author}{\bibinfo{person}{Apple} {and} \bibinfo{person}{Google}.} \bibinfo{year}{2025}\natexlab{}.
\newblock \bibinfo{booktitle}{\emph{Privacy-Preserving Contact Tracing}}.
\newblock
\urldef\tempurl%
\url{https://covid19.apple.com/contacttracing}
\showURL{%
\tempurl}


\bibitem[Arrieta-Ibarra et~al\mbox{.}(2018)]%
        {arrieta2018should}
\bibfield{author}{\bibinfo{person}{Imanol Arrieta-Ibarra}, \bibinfo{person}{Leonard Goff}, \bibinfo{person}{Diego Jim{\'e}nez-Hern{\'a}ndez}, \bibinfo{person}{Jaron Lanier}, {and} \bibinfo{person}{E~Glen Weyl}.} \bibinfo{year}{2018}\natexlab{}.
\newblock \showarticletitle{Should we treat data as labor? Moving beyond “free”}. In \bibinfo{booktitle}{\emph{aea Papers and Proceedings}}, Vol.~\bibinfo{volume}{108}. American Economic Association 2014 Broadway, Suite 305, Nashville, TN 37203, \bibinfo{pages}{38--42}.
\newblock


\bibitem[Assistant(2025)]%
        {Home-Assistant}
\bibfield{author}{\bibinfo{person}{Home Assistant}.} \bibinfo{year}{2025}\natexlab{}.
\newblock \bibinfo{booktitle}{\emph{Home Assistant for Apple Platforms}}.
\newblock
\urldef\tempurl%
\url{https://github.com/home-assistant/iOS}
\showURL{%
\tempurl}


\bibitem[Bluesky(2025)]%
        {Bluesky}
\bibfield{author}{\bibinfo{person}{Bluesky}.} \bibinfo{year}{2025}\natexlab{}.
\newblock \bibinfo{booktitle}{\emph{Bluesky Social App}}.
\newblock
\urldef\tempurl%
\url{https://github.com/bluesky-social/social-app}
\showURL{%
\tempurl}


\bibitem[Bravo-Lillo et~al\mbox{.}(2014)]%
        {bravo2014harder}
\bibfield{author}{\bibinfo{person}{Cristian Bravo-Lillo}, \bibinfo{person}{Lorrie Cranor}, \bibinfo{person}{Saranga Komanduri}, \bibinfo{person}{Stuart Schechter}, {and} \bibinfo{person}{Manya Sleeper}.} \bibinfo{year}{2014}\natexlab{}.
\newblock \showarticletitle{Harder to ignore? revisiting $\{$Pop-Up$\}$ fatigue and approaches to prevent it}. In \bibinfo{booktitle}{\emph{10th Symposium On Usable Privacy and Security (SOUPS 2014)}}. \bibinfo{pages}{105--111}.
\newblock


\bibitem[Brodie et~al\mbox{.}(2005)]%
        {brodie2005usable}
\bibfield{author}{\bibinfo{person}{Carolyn Brodie}, \bibinfo{person}{Clare-Marie Karat}, \bibinfo{person}{John Karat}, {and} \bibinfo{person}{Jinjuan Feng}.} \bibinfo{year}{2005}\natexlab{}.
\newblock \showarticletitle{Usable security and privacy: a case study of developing privacy management tools}. In \bibinfo{booktitle}{\emph{Proceedings of the 2005 symposium on Usable privacy and security}}. \bibinfo{pages}{35--43}.
\newblock


\bibitem[Byun and Li(2008)]%
        {byun2008purpose}
\bibfield{author}{\bibinfo{person}{Ji-Won Byun} {and} \bibinfo{person}{Ninghui Li}.} \bibinfo{year}{2008}\natexlab{}.
\newblock \showarticletitle{Purpose based access control for privacy protection in relational database systems}.
\newblock \bibinfo{journal}{\emph{The VLDB Journal}} \bibinfo{volume}{17}, \bibinfo{number}{4} (\bibinfo{year}{2008}), \bibinfo{pages}{603--619}.
\newblock


\bibitem[Canetti et~al\mbox{.}(1996)]%
        {canetti1996adaptively}
\bibfield{author}{\bibinfo{person}{Ran Canetti}, \bibinfo{person}{Uri Feige}, \bibinfo{person}{Oded Goldreich}, {and} \bibinfo{person}{Moni Naor}.} \bibinfo{year}{1996}\natexlab{}.
\newblock \showarticletitle{Adaptively secure multi-party computation}. In \bibinfo{booktitle}{\emph{Proceedings of the twenty-eighth annual ACM symposium on Theory of computing}}. \bibinfo{pages}{639--648}.
\newblock


\bibitem[Carey et~al\mbox{.}(1995)]%
        {carey1995towards}
\bibfield{author}{\bibinfo{person}{Michael~J Carey}, \bibinfo{person}{Laura~M Haas}, \bibinfo{person}{Peter~M Schwarz}, \bibinfo{person}{Manish Arya}, \bibinfo{person}{William~F Cody}, \bibinfo{person}{Ronald Fagin}, \bibinfo{person}{Myron Flickner}, \bibinfo{person}{Allen~W Luniewski}, \bibinfo{person}{Wayne Niblack}, \bibinfo{person}{Dragutin Petkovic}, {et~al\mbox{.}}} \bibinfo{year}{1995}\natexlab{}.
\newblock \showarticletitle{Towards heterogeneous multimedia information systems: The Garlic approach}. In \bibinfo{booktitle}{\emph{Proceedings RIDE-DOM'95. Fifth International Workshop on Research Issues in Data Engineering-Distributed Object Management}}. IEEE, \bibinfo{pages}{124--131}.
\newblock


\bibitem[Choi et~al\mbox{.}(2018)]%
        {choi2018role}
\bibfield{author}{\bibinfo{person}{Hanbyul Choi}, \bibinfo{person}{Jonghwa Park}, {and} \bibinfo{person}{Yoonhyuk Jung}.} \bibinfo{year}{2018}\natexlab{}.
\newblock \showarticletitle{The role of privacy fatigue in online privacy behavior}.
\newblock \bibinfo{journal}{\emph{Computers in Human Behavior}}  \bibinfo{volume}{81} (\bibinfo{year}{2018}), \bibinfo{pages}{42--51}.
\newblock


\bibitem[Claesson and Bj{\o}rstad(2020)]%
        {claesson2020technical}
\bibfield{author}{\bibinfo{person}{Andreas Claesson} {and} \bibinfo{person}{Tor~E Bj{\o}rstad}.} \bibinfo{year}{2020}\natexlab{}.
\newblock \showarticletitle{Technical report:‘out of control’—a review of data sharing by popular mobile apps}.
\newblock \bibinfo{journal}{\emph{Report for the Norwegian Consumer Council}}  \bibinfo{volume}{14} (\bibinfo{year}{2020}).
\newblock


\bibitem[Collier and Burke(2022)]%
        {collier2022facebook}
\bibfield{author}{\bibinfo{person}{Kevin Collier} {and} \bibinfo{person}{Minyvonne Burke}.} \bibinfo{year}{2022}\natexlab{}.
\newblock \bibinfo{title}{Facebook turned over chat messages between mother and daughter now charged over abortion}.
\newblock
\newblock


\bibitem[Crain(2018)]%
        {crain2018limits}
\bibfield{author}{\bibinfo{person}{Matthew Crain}.} \bibinfo{year}{2018}\natexlab{}.
\newblock \showarticletitle{The limits of transparency: Data brokers and commodification}.
\newblock \bibinfo{journal}{\emph{new media \& society}} \bibinfo{volume}{20}, \bibinfo{number}{1} (\bibinfo{year}{2018}), \bibinfo{pages}{88--104}.
\newblock


\bibitem[Delacroix and Lawrence(2019)]%
        {delacroix2019bottom}
\bibfield{author}{\bibinfo{person}{Sylvie Delacroix} {and} \bibinfo{person}{Neil~D Lawrence}.} \bibinfo{year}{2019}\natexlab{}.
\newblock \showarticletitle{Bottom-up data trusts: Disturbing the ‘one size fits all’approach to data governance}.
\newblock \bibinfo{journal}{\emph{International data privacy law}} \bibinfo{volume}{9}, \bibinfo{number}{4} (\bibinfo{year}{2019}), \bibinfo{pages}{236--252}.
\newblock


\bibitem[Distler et~al\mbox{.}(2021)]%
        {distler2021systematic}
\bibfield{author}{\bibinfo{person}{Verena Distler}, \bibinfo{person}{Matthias Fassl}, \bibinfo{person}{Hana Habib}, \bibinfo{person}{Katharina Krombholz}, \bibinfo{person}{Gabriele Lenzini}, \bibinfo{person}{Carine Lallemand}, \bibinfo{person}{Lorrie~Faith Cranor}, {and} \bibinfo{person}{Vincent Koenig}.} \bibinfo{year}{2021}\natexlab{}.
\newblock \showarticletitle{A systematic literature review of empirical methods and risk representation in usable privacy and security research}.
\newblock \bibinfo{journal}{\emph{ACM Transactions on Computer-Human Interaction (TOCHI)}} \bibinfo{volume}{28}, \bibinfo{number}{6} (\bibinfo{year}{2021}), \bibinfo{pages}{1--50}.
\newblock


\bibitem[DuckDuckGo(2025)]%
        {DuckDuckGo}
\bibfield{author}{\bibinfo{person}{DuckDuckGo}.} \bibinfo{year}{2025}\natexlab{}.
\newblock \bibinfo{booktitle}{\emph{DuckDuckGo iOS}}.
\newblock
\urldef\tempurl%
\url{https://github.com/duckduckgo/ios}
\showURL{%
\tempurl}


\bibitem[Duesterwald et~al\mbox{.}(2025)]%
        {duesterwald2025can}
\bibfield{author}{\bibinfo{person}{Lea Duesterwald}, \bibinfo{person}{Ian Yang}, {and} \bibinfo{person}{Norman Sadeh}.} \bibinfo{year}{2025}\natexlab{}.
\newblock \showarticletitle{Can a cybersecurity question answering assistant help change user behavior? an in situ study}. Symposium on Usable Security and Privacy (USEC) 2025-NDSS Symposium.
\newblock


\bibitem[Ernstberger et~al\mbox{.}(2023)]%
        {ernstberger2023sok}
\bibfield{author}{\bibinfo{person}{Jens Ernstberger}, \bibinfo{person}{Jan Lauinger}, \bibinfo{person}{Fatima Elsheimy}, \bibinfo{person}{Liyi Zhou}, \bibinfo{person}{Sebastian Steinhorst}, \bibinfo{person}{Ran Canetti}, \bibinfo{person}{Andrew Miller}, \bibinfo{person}{Arthur Gervais}, {and} \bibinfo{person}{Dawn Song}.} \bibinfo{year}{2023}\natexlab{}.
\newblock \showarticletitle{Sok: data sovereignty}. In \bibinfo{booktitle}{\emph{2023 IEEE 8th European Symposium on Security and Privacy (EuroS\&P)}}. IEEE, \bibinfo{pages}{122--143}.
\newblock


\bibitem[{European Parliament} and {Council of the European Union}(2016)]%
        {gdpr}
\bibfield{author}{\bibinfo{person}{{European Parliament}} {and} \bibinfo{person}{{Council of the European Union}}.} \bibinfo{year}{2016}\natexlab{}.
\newblock \bibinfo{booktitle}{\emph{Regulation ({EU}) 2016/679 of the {European} {Parliament} and of the {Council}}}.
\newblock
\urldef\tempurl%
\url{https://eur-lex.europa.eu/legal-content/EN/TXT/PDF/?uri=CELEX:32016R0679}
\showURL{%
\tempurl}


\bibitem[Felt et~al\mbox{.}(2012)]%
        {felt2012android}
\bibfield{author}{\bibinfo{person}{Adrienne~Porter Felt}, \bibinfo{person}{Elizabeth Ha}, \bibinfo{person}{Serge Egelman}, \bibinfo{person}{Ariel Haney}, \bibinfo{person}{Erika Chin}, {and} \bibinfo{person}{David Wagner}.} \bibinfo{year}{2012}\natexlab{}.
\newblock \showarticletitle{Android permissions: User attention, comprehension, and behavior}. In \bibinfo{booktitle}{\emph{Proceedings of the eighth symposium on usable privacy and security}}. \bibinfo{pages}{1--14}.
\newblock


\bibitem[Feng et~al\mbox{.}(2024)]%
        {feng2024understanding}
\bibfield{author}{\bibinfo{person}{Yuanyuan Feng}, \bibinfo{person}{Abhilasha Ravichander}, \bibinfo{person}{Yaxing Yao}, \bibinfo{person}{Shikun Zhang}, {and} \bibinfo{person}{Rex Chen}.} \bibinfo{year}{2024}\natexlab{}.
\newblock \showarticletitle{Understanding how to inform blind and $\{$Low-Vision$\}$ users about data privacy through privacy question answering assistants}. In \bibinfo{booktitle}{\emph{33rd USENIX Security Symposium (USENIX Security 24)}}. \bibinfo{pages}{2065--2082}.
\newblock


\bibitem[Feng et~al\mbox{.}(2021)]%
        {feng2021design}
\bibfield{author}{\bibinfo{person}{Yuanyuan Feng}, \bibinfo{person}{Yaxing Yao}, {and} \bibinfo{person}{Norman Sadeh}.} \bibinfo{year}{2021}\natexlab{}.
\newblock \showarticletitle{A design space for privacy choices: Towards meaningful privacy control in the internet of things}. In \bibinfo{booktitle}{\emph{Proceedings of the 2021 CHI Conference on Human Factors in Computing Systems}}. \bibinfo{pages}{1--16}.
\newblock


\bibitem[Fisch et~al\mbox{.}(2017)]%
        {fisch2017iron}
\bibfield{author}{\bibinfo{person}{Ben Fisch}, \bibinfo{person}{Dhinakaran Vinayagamurthy}, \bibinfo{person}{Dan Boneh}, {and} \bibinfo{person}{Sergey Gorbunov}.} \bibinfo{year}{2017}\natexlab{}.
\newblock \showarticletitle{Iron: functional encryption using Intel SGX}. In \bibinfo{booktitle}{\emph{Proceedings of the 2017 ACM SIGSAC Conference on Computer and Communications Security}}. \bibinfo{pages}{765--782}.
\newblock


\bibitem[Fischer-H{\"u}bner and Karegar(2024)]%
        {fischer2024overview}
\bibfield{author}{\bibinfo{person}{Simone Fischer-H{\"u}bner} {and} \bibinfo{person}{Farzaneh Karegar}.} \bibinfo{year}{2024}\natexlab{}.
\newblock \showarticletitle{Overview of Usable Privacy Research: Major Themes and Research Directions}.
\newblock \bibinfo{journal}{\emph{The Curious Case of Usable Privacy: Challenges, Solutions, and Prospects}} (\bibinfo{year}{2024}), \bibinfo{pages}{43--102}.
\newblock


\bibitem[Gamiz~Ugarte et~al\mbox{.}(2025)]%
        {gamiz2025challenges}
\bibfield{author}{\bibinfo{person}{Idoia Gamiz~Ugarte}, \bibinfo{person}{Cristina Regueiro~Senderos}, \bibinfo{person}{{\'O}scar Lage~Serrano}, \bibinfo{person}{Eduardo Jacob~Taquet}, {and} \bibinfo{person}{Jasone Astorga~Burgo}.} \bibinfo{year}{2025}\natexlab{}.
\newblock \showarticletitle{Challenges and future research directions in secure multi-party computation for resource-constrained devices and large-scale computations}.
\newblock \bibinfo{journal}{\emph{International Journal of Information Security}}  \bibinfo{volume}{24} (\bibinfo{year}{2025}).
\newblock


\bibitem[Google(2025)]%
        {protobuf}
\bibfield{author}{\bibinfo{person}{Google}.} \bibinfo{year}{2025}\natexlab{}.
\newblock \bibinfo{booktitle}{\emph{Protocol Buffers}}.
\newblock
\urldef\tempurl%
\url{https://protobuf.dev/}
\showURL{%
\tempurl}


\bibitem[Habib and Cranor(2022)]%
        {habib2022evaluating}
\bibfield{author}{\bibinfo{person}{Hana Habib} {and} \bibinfo{person}{Lorrie~Faith Cranor}.} \bibinfo{year}{2022}\natexlab{}.
\newblock \showarticletitle{Evaluating the usability of privacy choice mechanisms}. In \bibinfo{booktitle}{\emph{Eighteenth Symposium on Usable Privacy and Security (SOUPS 2022)}}. \bibinfo{pages}{273--289}.
\newblock


\bibitem[Halevy(2001)]%
        {halevy2001answering}
\bibfield{author}{\bibinfo{person}{Alon~Y Halevy}.} \bibinfo{year}{2001}\natexlab{}.
\newblock \showarticletitle{Answering queries using views: A survey}.
\newblock \bibinfo{journal}{\emph{The VLDB Journal}} \bibinfo{volume}{10}, \bibinfo{number}{4} (\bibinfo{year}{2001}), \bibinfo{pages}{270--294}.
\newblock


\bibitem[Hellerstein and Stonebraker(1993)]%
        {hellerstein1993predicate}
\bibfield{author}{\bibinfo{person}{Joseph~M Hellerstein} {and} \bibinfo{person}{Michael Stonebraker}.} \bibinfo{year}{1993}\natexlab{}.
\newblock \showarticletitle{Predicate migration: Optimizing queries with expensive predicates}. In \bibinfo{booktitle}{\emph{Proceedings of the 1993 ACM SIGMOD international conference on Management of data}}. \bibinfo{pages}{267--276}.
\newblock


\bibitem[Hummel et~al\mbox{.}(2021)]%
        {hummel2021data}
\bibfield{author}{\bibinfo{person}{Patrik Hummel}, \bibinfo{person}{Matthias Braun}, \bibinfo{person}{Max Tretter}, {and} \bibinfo{person}{Peter Dabrock}.} \bibinfo{year}{2021}\natexlab{}.
\newblock \showarticletitle{Data sovereignty: A review}.
\newblock \bibinfo{journal}{\emph{Big Data \& Society}} \bibinfo{volume}{8}, \bibinfo{number}{1} (\bibinfo{year}{2021}), \bibinfo{pages}{2053951720982012}.
\newblock


\bibitem[Hunt et~al\mbox{.}(2020)]%
        {hunt2020telekine}
\bibfield{author}{\bibinfo{person}{Tyler Hunt}, \bibinfo{person}{Zhipeng Jia}, \bibinfo{person}{Vance Miller}, \bibinfo{person}{Ariel Szekely}, \bibinfo{person}{Yige Hu}, \bibinfo{person}{Christopher~J Rossbach}, {and} \bibinfo{person}{Emmett Witchel}.} \bibinfo{year}{2020}\natexlab{}.
\newblock \showarticletitle{Telekine: Secure computing with cloud $\{$GPUs$\}$}. In \bibinfo{booktitle}{\emph{17th USENIX Symposium on Networked Systems Design and Implementation (NSDI 20)}}. \bibinfo{pages}{817--833}.
\newblock


\bibitem[Institute(2021)]%
        {ada2021exploring}
\bibfield{author}{\bibinfo{person}{Ada~Lovelace Institute}.} \bibinfo{year}{2021}\natexlab{}.
\newblock \showarticletitle{Exploring legal mechanisms for data stewardship}.
\newblock \bibinfo{journal}{\emph{Ada Lovelace Institute and UK AI Council}} (\bibinfo{year}{2021}).
\newblock


\bibitem[Iravantchi et~al\mbox{.}(2025)]%
        {iravantchi2025sok}
\bibfield{author}{\bibinfo{person}{Yasha Iravantchi}, \bibinfo{person}{Pardis Emami-Naeini}, {and} \bibinfo{person}{Alanson Sample}.} \bibinfo{year}{2025}\natexlab{}.
\newblock \showarticletitle{Sok:(un) usable privacy: the lack of overlap between privacy-aware sensing and usable privacy research}.
\newblock \bibinfo{journal}{\emph{Proceedings on Privacy Enhancing Technologies}} (\bibinfo{year}{2025}).
\newblock


\bibitem[Ireland et~al\mbox{.}(2009)]%
        {ireland2009classification}
\bibfield{author}{\bibinfo{person}{Christopher Ireland}, \bibinfo{person}{David Bowers}, \bibinfo{person}{Michael Newton}, {and} \bibinfo{person}{Kevin Waugh}.} \bibinfo{year}{2009}\natexlab{}.
\newblock \showarticletitle{A classification of object-relational impedance mismatch}. In \bibinfo{booktitle}{\emph{2009 First International Confernce on Advances in Databases, Knowledge, and Data Applications}}. IEEE, \bibinfo{pages}{36--43}.
\newblock


\bibitem[Koch et~al\mbox{.}(2025)]%
        {koch2025impact}
\bibfield{author}{\bibinfo{person}{Simon Koch}, \bibinfo{person}{Manuel Karl}, \bibinfo{person}{Robin Kirchner}, \bibinfo{person}{Malte Wessels}, \bibinfo{person}{Anne Paschke}, {and} \bibinfo{person}{Martin Johns}.} \bibinfo{year}{2025}\natexlab{}.
\newblock \showarticletitle{The Impact of Default Mobile SDK Usage on Privacy and Data Protection}.
\newblock \bibinfo{journal}{\emph{Proceedings on Privacy Enhancing Technologies}} (\bibinfo{year}{2025}).
\newblock


\bibitem[Kumar et~al\mbox{.}(2013)]%
        {kumar2013survey}
\bibfield{author}{\bibinfo{person}{Karthik Kumar}, \bibinfo{person}{Jibang Liu}, \bibinfo{person}{Yung-Hsiang Lu}, {and} \bibinfo{person}{Bharat Bhargava}.} \bibinfo{year}{2013}\natexlab{}.
\newblock \showarticletitle{A survey of computation offloading for mobile systems}.
\newblock \bibinfo{journal}{\emph{Mobile networks and Applications}}  \bibinfo{volume}{18} (\bibinfo{year}{2013}), \bibinfo{pages}{129--140}.
\newblock


\bibitem[Lauf et~al\mbox{.}(2022)]%
        {lauf2022linking}
\bibfield{author}{\bibinfo{person}{Florian Lauf}, \bibinfo{person}{Simon Scheider}, \bibinfo{person}{Jan Bartsch}, \bibinfo{person}{Philipp Herrmann}, \bibinfo{person}{Marija Radic}, \bibinfo{person}{Marcel Rebbert}, \bibinfo{person}{Andr{\'e}~T Nemat}, \bibinfo{person}{Christoph Schlueter~Langdon}, \bibinfo{person}{Ralf Konrad}, \bibinfo{person}{Ali Sunyaev}, {et~al\mbox{.}}} \bibinfo{year}{2022}\natexlab{}.
\newblock \showarticletitle{Linking data sovereignty and data economy: arising areas of tension}.
\newblock  (\bibinfo{year}{2022}).
\newblock


\bibitem[Lentz et~al\mbox{.}(2018)]%
        {lentz2018secloak}
\bibfield{author}{\bibinfo{person}{Matthew Lentz}, \bibinfo{person}{Rijurekha Sen}, \bibinfo{person}{Peter Druschel}, {and} \bibinfo{person}{Bobby Bhattacharjee}.} \bibinfo{year}{2018}\natexlab{}.
\newblock \showarticletitle{Secloak: Arm trustzone-based mobile peripheral control}. In \bibinfo{booktitle}{\emph{Proceedings of the 16th Annual International Conference on Mobile Systems, Applications, and Services}}. \bibinfo{pages}{1--13}.
\newblock


\bibitem[Madden et~al\mbox{.}(2005)]%
        {madden2005tinydb}
\bibfield{author}{\bibinfo{person}{Samuel~R Madden}, \bibinfo{person}{Michael~J Franklin}, \bibinfo{person}{Joseph~M Hellerstein}, {and} \bibinfo{person}{Wei Hong}.} \bibinfo{year}{2005}\natexlab{}.
\newblock \showarticletitle{TinyDB: an acquisitional query processing system for sensor networks}.
\newblock \bibinfo{journal}{\emph{ACM Transactions on database systems (TODS)}} \bibinfo{volume}{30}, \bibinfo{number}{1} (\bibinfo{year}{2005}), \bibinfo{pages}{122--173}.
\newblock


\bibitem[{Matthew Forsythe, Director Product Management, Android App Safety}(2026)]%
        {android-sideloading}
\bibfield{author}{\bibinfo{person}{{Matthew Forsythe, Director Product Management, Android App Safety}}.} \bibinfo{year}{2026}\natexlab{}.
\newblock \bibinfo{booktitle}{\emph{Android developer verification: Balancing openness and choice with safety}}.
\newblock
\urldef\tempurl%
\url{https://android-developers.googleblog.com/2026/03/android-developer-verification.html}
\showURL{%
\tempurl}


\bibitem[Nextcloud(2025)]%
        {Nextcloud}
\bibfield{author}{\bibinfo{person}{Nextcloud}.} \bibinfo{year}{2025}\natexlab{}.
\newblock \bibinfo{booktitle}{\emph{Nextcloud iOS ap}}.
\newblock
\urldef\tempurl%
\url{https://github.com/nextcloud/ios}
\showURL{%
\tempurl}


\bibitem[Nissenbaum(2004)]%
        {nissenbaum2004privacy}
\bibfield{author}{\bibinfo{person}{Helen Nissenbaum}.} \bibinfo{year}{2004}\natexlab{}.
\newblock \showarticletitle{Privacy as contextual integrity}.
\newblock \bibinfo{journal}{\emph{Wash. L. Rev.}}  \bibinfo{volume}{79} (\bibinfo{year}{2004}), \bibinfo{pages}{119}.
\newblock


\bibitem[Obar and Oeldorf-Hirsch(2020)]%
        {obar2020biggest}
\bibfield{author}{\bibinfo{person}{Jonathan~A Obar} {and} \bibinfo{person}{Anne Oeldorf-Hirsch}.} \bibinfo{year}{2020}\natexlab{}.
\newblock \showarticletitle{The biggest lie on the internet: Ignoring the privacy policies and terms of service policies of social networking services}.
\newblock \bibinfo{journal}{\emph{Information, Communication \& Society}} \bibinfo{volume}{23}, \bibinfo{number}{1} (\bibinfo{year}{2020}), \bibinfo{pages}{128--147}.
\newblock


\bibitem[OpenWeatherMap(2025)]%
        {openweathermap}
\bibfield{author}{\bibinfo{person}{OpenWeatherMap}.} \bibinfo{year}{2025}\natexlab{}.
\newblock \bibinfo{booktitle}{\emph{OpenWeatherMap API for current weather data}}.
\newblock
\urldef\tempurl%
\url{https://openweathermap.org/current}
\showURL{%
\tempurl}


\bibitem[Pan et~al\mbox{.}(2024)]%
        {pan2024new}
\bibfield{author}{\bibinfo{person}{Shidong Pan}, \bibinfo{person}{Zhen Tao}, \bibinfo{person}{Thong Hoang}, \bibinfo{person}{Dawen Zhang}, \bibinfo{person}{Tianshi Li}, \bibinfo{person}{Zhenchang Xing}, \bibinfo{person}{Xiwei Xu}, \bibinfo{person}{Mark Staples}, \bibinfo{person}{Thierry Rakotoarivelo}, {and} \bibinfo{person}{David Lo}.} \bibinfo{year}{2024}\natexlab{}.
\newblock \showarticletitle{A $\{$NEW$\}$$\{$HOPE$\}$: Contextual privacy policies for mobile applications and an approach toward automated generation}. In \bibinfo{booktitle}{\emph{33rd USENIX Security Symposium (USENIX Security 24)}}. \bibinfo{pages}{5699--5716}.
\newblock


\bibitem[Pielot et~al\mbox{.}(2018)]%
        {pielot2018dismissed}
\bibfield{author}{\bibinfo{person}{Martin Pielot}, \bibinfo{person}{Amalia Vradi}, {and} \bibinfo{person}{Souneil Park}.} \bibinfo{year}{2018}\natexlab{}.
\newblock \showarticletitle{Dismissed! a detailed exploration of how mobile phone users handle push notifications}. In \bibinfo{booktitle}{\emph{Proceedings of the 20th international conference on human-computer interaction with mobile devices and services}}. \bibinfo{pages}{1--11}.
\newblock


\bibitem[Pol{\v{c}}{\'a}k and Svantesson(2017)]%
        {polvcak2017information}
\bibfield{author}{\bibinfo{person}{Radim Pol{\v{c}}{\'a}k} {and} \bibinfo{person}{Dan Jerker~B Svantesson}.} \bibinfo{year}{2017}\natexlab{}.
\newblock \bibinfo{booktitle}{\emph{Information sovereignty: data privacy, sovereign powers and the rule of law}}.
\newblock \bibinfo{publisher}{Edward Elgar Publishing}.
\newblock


\bibitem[Popa(2024)]%
        {popa2024confidential}
\bibfield{author}{\bibinfo{person}{Raluca~Ada Popa}.} \bibinfo{year}{2024}\natexlab{}.
\newblock \showarticletitle{Confidential Computing or Cryptographic Computing? Tradeoffs between cryptography and hardware enclaves}.
\newblock \bibinfo{journal}{\emph{Queue}} \bibinfo{volume}{22}, \bibinfo{number}{2} (\bibinfo{year}{2024}), \bibinfo{pages}{108--132}.
\newblock


\bibitem[Posner and Weyl(2018)]%
        {posner2018radical}
\bibfield{author}{\bibinfo{person}{Eric Posner} {and} \bibinfo{person}{Eric Weyl}.} \bibinfo{year}{2018}\natexlab{}.
\newblock \bibinfo{booktitle}{\emph{Radical markets: Uprooting capitalism and democracy for a just society}}.
\newblock \bibinfo{publisher}{Princeton University Press}.
\newblock


\bibitem[PostgreSQ(2025)]%
        {postgres-fdw}
\bibfield{author}{\bibinfo{person}{PostgreSQ}.} \bibinfo{year}{2025}\natexlab{}.
\newblock \bibinfo{booktitle}{\emph{Foreign Data}}.
\newblock
\urldef\tempurl%
\url{https://www.postgresql.org/docs/current/ddl-foreign-data.html}
\showURL{%
\tempurl}


\bibitem[researc(2020)]%
        {mozilla-shifting-power}
\bibfield{author}{\bibinfo{person}{Mozilla researc}.} \bibinfo{year}{2020}\natexlab{}.
\newblock \bibinfo{booktitle}{\emph{What Does it Mean? | Shifting Power Through Data Governance}}.
\newblock
\urldef\tempurl%
\url{https://foundation.mozilla.org/en/data-futures-lab/data-for-empowerment/shifting-power-through-data-governance/}
\showURL{%
\tempurl}


\bibitem[Rodriguez et~al\mbox{.}(2025)]%
        {rodriguez2025privacy}
\bibfield{author}{\bibinfo{person}{David Rodriguez}, \bibinfo{person}{Joseph~A Calandrino}, \bibinfo{person}{Jose~M Del~Alamo}, {and} \bibinfo{person}{Norman Sadeh}.} \bibinfo{year}{2025}\natexlab{}.
\newblock \showarticletitle{Privacy settings of third-party libraries in android apps: A study of facebook sdks}.
\newblock \bibinfo{journal}{\emph{Proceedings on Privacy Enhancing Technologies}} (\bibinfo{year}{2025}).
\newblock


\bibitem[Rodriguez et~al\mbox{.}(2024)]%
        {rodriguez2024sharing}
\bibfield{author}{\bibinfo{person}{David Rodriguez}, \bibinfo{person}{Jose~M Del~Alamo}, \bibinfo{person}{Celia Fern{\'a}ndez-Aller}, {and} \bibinfo{person}{Norman Sadeh}.} \bibinfo{year}{2024}\natexlab{}.
\newblock \showarticletitle{Sharing is not always caring: Delving into personal data transfer compliance in android apps}.
\newblock \bibinfo{journal}{\emph{IEEE Access}}  \bibinfo{volume}{12} (\bibinfo{year}{2024}), \bibinfo{pages}{5256--5269}.
\newblock


\bibitem[Ryan et~al\mbox{.}(2024)]%
        {ryan2024will}
\bibfield{author}{\bibinfo{person}{Mark Ryan}, \bibinfo{person}{Paula G{\"u}rtler}, {and} \bibinfo{person}{Artur Bogucki}.} \bibinfo{year}{2024}\natexlab{}.
\newblock \showarticletitle{Will the real data sovereign please stand up? An EU policy response to sovereignty in data spaces}.
\newblock \bibinfo{journal}{\emph{International Journal of Law and Information Technology}}  \bibinfo{volume}{32} (\bibinfo{year}{2024}), \bibinfo{pages}{eaae006}.
\newblock


\bibitem[Sabt et~al\mbox{.}(2015)]%
        {sabt2015trusted}
\bibfield{author}{\bibinfo{person}{Mohamed Sabt}, \bibinfo{person}{Mohammed Achemlal}, {and} \bibinfo{person}{Abdelmadjid Bouabdallah}.} \bibinfo{year}{2015}\natexlab{}.
\newblock \showarticletitle{Trusted execution environment: What it is, and what it is not}. In \bibinfo{booktitle}{\emph{2015 IEEE Trustcom/BigDataSE/Ispa}}, Vol.~\bibinfo{volume}{1}. IEEE, \bibinfo{pages}{57--64}.
\newblock


\bibitem[Schaub et~al\mbox{.}(2015)]%
        {schaub2015design}
\bibfield{author}{\bibinfo{person}{Florian Schaub}, \bibinfo{person}{Rebecca Balebako}, \bibinfo{person}{Adam~L Durity}, {and} \bibinfo{person}{Lorrie~Faith Cranor}.} \bibinfo{year}{2015}\natexlab{}.
\newblock \showarticletitle{A design space for effective privacy notices}. In \bibinfo{booktitle}{\emph{Eleventh symposium on usable privacy and security (SOUPS 2015)}}. \bibinfo{pages}{1--17}.
\newblock


\bibitem[Service(2025)]%
        {covid-alert}
\bibfield{author}{\bibinfo{person}{Canadian~Digital Service}.} \bibinfo{year}{2025}\natexlab{}.
\newblock \bibinfo{booktitle}{\emph{COVID Alert Mobile App}}.
\newblock
\urldef\tempurl%
\url{https://github.com/cds-snc/covid-alert-app}
\showURL{%
\tempurl}


\bibitem[Signal(2022)]%
        {signal-enclave}
\bibfield{author}{\bibinfo{person}{Signal}.} \bibinfo{year}{2022}\natexlab{}.
\newblock \bibinfo{booktitle}{\emph{Technology Deep Dive: Building a Faster ORAM Layer for Enclaves}}.
\newblock
\urldef\tempurl%
\url{https://signal.org/blog/building-faster-oram/}
\showURL{%
\tempurl}


\bibitem[Signal(2025)]%
        {Signal}
\bibfield{author}{\bibinfo{person}{Signal}.} \bibinfo{year}{2025}\natexlab{}.
\newblock \bibinfo{booktitle}{\emph{Signal iOS}}.
\newblock
\urldef\tempurl%
\url{https://github.com/signalapp/Signal-iOS}
\showURL{%
\tempurl}


\bibitem[Simplenote(2025)]%
        {Simplenote}
\bibfield{author}{\bibinfo{person}{Simplenote}.} \bibinfo{year}{2025}\natexlab{}.
\newblock \bibinfo{booktitle}{\emph{Simplenote for iOS}}.
\newblock
\urldef\tempurl%
\url{https://github.com/automattic/simplenote-ios}
\showURL{%
\tempurl}


\bibitem[Spinks(2019)]%
        {spinks2019contemporary}
\bibfield{author}{\bibinfo{person}{Chandler~Nicholle Spinks}.} \bibinfo{year}{2019}\natexlab{}.
\newblock \showarticletitle{Contemporary housing discrimination: Facebook, targeted advertising, and the Fair Housing Act}.
\newblock \bibinfo{journal}{\emph{Hous. L. Rev.}}  \bibinfo{volume}{57} (\bibinfo{year}{2019}), \bibinfo{pages}{925}.
\newblock


\bibitem[SQLite(2025a)]%
        {sqlite3_index_info}
\bibfield{author}{\bibinfo{person}{SQLite}.} \bibinfo{year}{2025}\natexlab{a}.
\newblock \bibinfo{booktitle}{\emph{Virtual Table Indexing Information}}.
\newblock
\urldef\tempurl%
\url{https://www.sqlite.org/c3ref/index_info.html}
\showURL{%
\tempurl}


\bibitem[SQLite(2025b)]%
        {sqlite-vtab}
\bibfield{author}{\bibinfo{person}{SQLite}.} \bibinfo{year}{2025}\natexlab{b}.
\newblock \bibinfo{booktitle}{\emph{The Virtual Table Mechanism Of SQLite}}.
\newblock
\urldef\tempurl%
\url{https://www.sqlite.org/vtab.html}
\showURL{%
\tempurl}


\bibitem[SQLite(2025c)]%
        {sqlite3_module}
\bibfield{author}{\bibinfo{person}{SQLite}.} \bibinfo{year}{2025}\natexlab{c}.
\newblock \bibinfo{booktitle}{\emph{Virtual Table Object}}.
\newblock
\urldef\tempurl%
\url{https://www.sqlite.org/c3ref/module.html}
\showURL{%
\tempurl}


\bibitem[Sun et~al\mbox{.}(2023)]%
        {sun2023shadownet}
\bibfield{author}{\bibinfo{person}{Zhichuang Sun}, \bibinfo{person}{Ruimin Sun}, \bibinfo{person}{Changming Liu}, \bibinfo{person}{Amrita~Roy Chowdhury}, \bibinfo{person}{Long Lu}, {and} \bibinfo{person}{Somesh Jha}.} \bibinfo{year}{2023}\natexlab{}.
\newblock \showarticletitle{Shadownet: A secure and efficient on-device model inference system for convolutional neural networks}. In \bibinfo{booktitle}{\emph{2023 IEEE Symposium on Security and Privacy (SP)}}. IEEE, \bibinfo{pages}{1596--1612}.
\newblock


\bibitem[Tan et~al\mbox{.}(2014)]%
        {tan2014effect}
\bibfield{author}{\bibinfo{person}{Joshua Tan}, \bibinfo{person}{Khanh Nguyen}, \bibinfo{person}{Michael Theodorides}, \bibinfo{person}{Heidi Negr{\'o}n-Arroyo}, \bibinfo{person}{Christopher Thompson}, \bibinfo{person}{Serge Egelman}, {and} \bibinfo{person}{David Wagner}.} \bibinfo{year}{2014}\natexlab{}.
\newblock \showarticletitle{The effect of developer-specified explanations for permission requests on smartphone user behavior}. In \bibinfo{booktitle}{\emph{Proceedings of the SIGCHI Conference on Human Factors in Computing Systems}}. \bibinfo{pages}{91--100}.
\newblock


\bibitem[Tramer and Boneh(2018)]%
        {tramer2018slalom}
\bibfield{author}{\bibinfo{person}{Florian Tramer} {and} \bibinfo{person}{Dan Boneh}.} \bibinfo{year}{2018}\natexlab{}.
\newblock \showarticletitle{Slalom: Fast, verifiable and private execution of neural networks in trusted hardware}.
\newblock \bibinfo{journal}{\emph{arXiv preprint arXiv:1806.03287}} (\bibinfo{year}{2018}).
\newblock


\bibitem[VLC(2025)]%
        {VLC}
\bibfield{author}{\bibinfo{person}{VLC}.} \bibinfo{year}{2025}\natexlab{}.
\newblock \bibinfo{booktitle}{\emph{VLC media player}}.
\newblock
\urldef\tempurl%
\url{https://github.com/videolan/vlc}
\showURL{%
\tempurl}


\bibitem[Wang et~al\mbox{.}(2016)]%
        {wang2016sieve}
\bibfield{author}{\bibinfo{person}{Frank Wang}, \bibinfo{person}{James Mickens}, \bibinfo{person}{Nickolai Zeldovich}, {and} \bibinfo{person}{Vinod Vaikuntanathan}.} \bibinfo{year}{2016}\natexlab{}.
\newblock \showarticletitle{Sieve: Cryptographically enforced access control for user data in untrusted clouds}. In \bibinfo{booktitle}{\emph{13th USENIX Symposium on Networked Systems Design and Implementation (NSDI 16)}}. \bibinfo{pages}{611--626}.
\newblock


\bibitem[Wikipedia(2025)]%
        {Wikipedia}
\bibfield{author}{\bibinfo{person}{Wikipedia}.} \bibinfo{year}{2025}\natexlab{}.
\newblock \bibinfo{booktitle}{\emph{Wikipedia iOS}}.
\newblock
\urldef\tempurl%
\url{https://github.com/wikimedia/wikipedia-ios}
\showURL{%
\tempurl}


\bibitem[WordPress(2025)]%
        {WordPress}
\bibfield{author}{\bibinfo{person}{WordPress}.} \bibinfo{year}{2025}\natexlab{}.
\newblock \bibinfo{booktitle}{\emph{WordPress for iOS}}.
\newblock
\urldef\tempurl%
\url{https://github.com/wordpress-mobile/WordPress-iOS}
\showURL{%
\tempurl}


\bibitem[Xia et~al\mbox{.}(2023)]%
        {xia2023data}
\bibfield{author}{\bibinfo{person}{Siyuan Xia}, \bibinfo{person}{Zhiru Zhu}, \bibinfo{person}{Chris Zhu}, \bibinfo{person}{Jinjin Zhao}, \bibinfo{person}{Kyle Chard}, \bibinfo{person}{Aaron~J Elmore}, \bibinfo{person}{Ian Foster}, \bibinfo{person}{Michael Franklin}, \bibinfo{person}{Sanjay Krishnan}, {and} \bibinfo{person}{Raul~Castro Fernandez}.} \bibinfo{year}{2023}\natexlab{}.
\newblock \showarticletitle{Data station: delegated, trustworthy, and auditable computation to enable data-sharing consortia with a data escrow}.
\newblock \bibinfo{journal}{\emph{arXiv preprint arXiv:2305.03842}} (\bibinfo{year}{2023}).
\newblock


\bibitem[Xia et~al\mbox{.}(2015)]%
        {xia2015tinman}
\bibfield{author}{\bibinfo{person}{Yubin Xia}, \bibinfo{person}{Yutao Liu}, \bibinfo{person}{Cheng Tan}, \bibinfo{person}{Mingyang Ma}, \bibinfo{person}{Haibing Guan}, \bibinfo{person}{Binyu Zang}, {and} \bibinfo{person}{Haibo Chen}.} \bibinfo{year}{2015}\natexlab{}.
\newblock \showarticletitle{TinMan: Eliminating confidential mobile data exposure with security oriented offloading}. In \bibinfo{booktitle}{\emph{Proceedings of the Tenth European Conference on Computer Systems}}. \bibinfo{pages}{1--16}.
\newblock


\bibitem[Xiao et~al\mbox{.}(2023)]%
        {xiao2023lalaine}
\bibfield{author}{\bibinfo{person}{Yue Xiao}, \bibinfo{person}{Zhengyi Li}, \bibinfo{person}{Yue Qin}, \bibinfo{person}{Xiaolong Bai}, \bibinfo{person}{Jiale Guan}, \bibinfo{person}{Xiaojing Liao}, {and} \bibinfo{person}{Luyi Xing}.} \bibinfo{year}{2023}\natexlab{}.
\newblock \showarticletitle{Lalaine: Measuring and Characterizing $\{$Non-Compliance$\}$ of Apple Privacy Labels}. In \bibinfo{booktitle}{\emph{32nd USENIX Security Symposium (USENIX Security 23)}}. \bibinfo{pages}{1091--1108}.
\newblock


\bibitem[Yang(2024)]%
        {yang2024creative}
\bibfield{author}{\bibinfo{person}{Angela Yang}.} \bibinfo{year}{2024}\natexlab{}.
\newblock \bibinfo{title}{More than 11,000 creatives condemn unauthorized use of content for AI development}.
\newblock
\newblock


\bibitem[Zang et~al\mbox{.}(2015)]%
        {zang2015knows}
\bibfield{author}{\bibinfo{person}{Jinyan Zang}, \bibinfo{person}{Krysta Dummit}, \bibinfo{person}{James Graves}, \bibinfo{person}{Paul Lisker}, {and} \bibinfo{person}{Latanya Sweeney}.} \bibinfo{year}{2015}\natexlab{}.
\newblock \showarticletitle{Who knows what about me? A survey of behind the scenes personal data sharing to third parties by mobile apps}.
\newblock \bibinfo{journal}{\emph{Technology Science}}  \bibinfo{volume}{30} (\bibinfo{year}{2015}), \bibinfo{pages}{2--1}.
\newblock


\bibitem[Zhang et~al\mbox{.}(2022)]%
        {zhang2022usable}
\bibfield{author}{\bibinfo{person}{Shikun Zhang}, \bibinfo{person}{Yuanyuan Feng}, \bibinfo{person}{Yaxing Yao}, \bibinfo{person}{Lorrie~Faith Cranor}, {and} \bibinfo{person}{Norman Sadeh}.} \bibinfo{year}{2022}\natexlab{}.
\newblock \showarticletitle{How usable are ios app privacy labels?}
\newblock \bibinfo{journal}{\emph{UMBC Faculty Collection}} (\bibinfo{year}{2022}).
\newblock


\bibitem[Zhang et~al\mbox{.}(2024b)]%
        {zhang2024exploring}
\bibfield{author}{\bibinfo{person}{Shikun Zhang}, \bibinfo{person}{Lily Klucinec}, \bibinfo{person}{Kyerra Norton}, \bibinfo{person}{Norman Sadeh}, {and} \bibinfo{person}{Lorrie~Faith Cranor}.} \bibinfo{year}{2024}\natexlab{b}.
\newblock \showarticletitle{Exploring $\{$Expandable-Grid$\}$ Designs to Make $\{$iOS$\}$ App Privacy Labels More Usable}. In \bibinfo{booktitle}{\emph{Twentieth Symposium on Usable Privacy and Security (SOUPS 2024)}}. \bibinfo{pages}{139--157}.
\newblock


\bibitem[Zhang and Sadeh(2023)]%
        {zhang2023privacy}
\bibfield{author}{\bibinfo{person}{Shikun Zhang} {and} \bibinfo{person}{Norman Sadeh}.} \bibinfo{year}{2023}\natexlab{}.
\newblock \showarticletitle{Do privacy labels answer users’ privacy questions}. In \bibinfo{booktitle}{\emph{Network and Distributed System Security Symposium}}.
\newblock


\bibitem[Zhang et~al\mbox{.}(2024a)]%
        {zhang2024navigating}
\bibfield{author}{\bibinfo{person}{Yifan Zhang}, \bibinfo{person}{Zhaojie Hu}, \bibinfo{person}{Xueqiang Wang}, \bibinfo{person}{Yuhui Hong}, \bibinfo{person}{Yuhong Nan}, \bibinfo{person}{XiaoFeng Wang}, \bibinfo{person}{Jiatao Cheng}, {and} \bibinfo{person}{Luyi Xing}.} \bibinfo{year}{2024}\natexlab{a}.
\newblock \showarticletitle{Navigating the Privacy Compliance Maze: Understanding Risks with $\{$Privacy-Configurable$\}$ Mobile $\{$SDKs$\}$}. In \bibinfo{booktitle}{\emph{33rd USENIX Security Symposium (USENIX Security 24)}}. \bibinfo{pages}{6543--6560}.
\newblock


\bibitem[Zimperium(2025)]%
        {mobile-threat-report}
\bibfield{author}{\bibinfo{person}{Zimperium}.} \bibinfo{year}{2025}\natexlab{}.
\newblock \bibinfo{booktitle}{\emph{2025 Zimperium Global Mobile Threat Report}}.
\newblock
\urldef\tempurl%
\url{https://lp.zimperium.com/hubfs/Reports/2025%20Global%20Mobile%20Threat%20Report.pdf}
\showURL{%
\tempurl}


\bibitem[Zuo et~al\mbox{.}(2019)]%
        {zuo2019does}
\bibfield{author}{\bibinfo{person}{Chaoshun Zuo}, \bibinfo{person}{Zhiqiang Lin}, {and} \bibinfo{person}{Yinqian Zhang}.} \bibinfo{year}{2019}\natexlab{}.
\newblock \showarticletitle{Why does your data leak? uncovering the data leakage in cloud from mobile apps}. In \bibinfo{booktitle}{\emph{2019 IEEE Symposium on Security and Privacy (SP)}}. IEEE, \bibinfo{pages}{1296--1310}.
\newblock


\end{thebibliography}


\end{document}